\providecommand{\paper}[1]{#1}

\paper{
\newcommand{\thesis}[1]{}
\newcommand{\paperchap}{paper\xspace}

\documentclass[a4paper]{article}
\usepackage{amsmath}
\usepackage{amssymb}
\usepackage{graphicx}
\usepackage{xy}
\usepackage[thmmarks]{ntheorem}
\usepackage{xspace}

\frenchspacing \setlength{\parindent}{0pt} \setlength{\parskip}{1ex
plus 0.5ex minus 0.2ex}

\title{The Fibers and Range of Reduction Graphs in Ciliates}
\author{Robert Brijder and Hendrik Jan Hoogeboom\\
\mbox{ }\\
Leiden Institute of Advanced Computer Science, Universiteit Leiden,\\
Niels Bohrweg 1, 2333 CA Leiden, The Netherlands,\\
\texttt{rbrijder@liacs.nl}}

\def\blfootnote{\xdef\@thefnmark{}\@footnotetext}
\long\def\symbolfootnote[#1]#2{\begingroup%
\def\thefootnote{\fnsymbol{footnote}}\footnote[#1]{#2}\endgroup}

\xyoption{all}

\theoremstyle{break} \theorembodyfont{\upshape}

\theoremsymbol{}
\newtheorem{Theorem}{Theorem}
\newtheorem{Lemma}[Theorem]{Lemma}
\newtheorem{Corollary}[Theorem]{Corollary}
\newtheorem{Example}{Example}

\theoremsymbol{\rule{1ex}{1ex}}
\newtheorem{Definition}[Theorem]{Definition}

\theoremstyle{nonumberbreak}
\newtheorem{Proof}{Proof}

\theoremsymbol{}
\newtheorem{Remark}{Remark}

\begin{document}
\bibliographystyle{plain}

\maketitle

\newcommand{\rf}{r \! f}
\newcommand{\pset}[1]{{\mathbf #1}}
\newcommand{\merge}{\mathrm{merge}}
\newcommand{\RGVertL}[1]{I_{#1}}
\newcommand{\RGVertR}[1]{I'_{#1}}
\newcommand{\medge}{\ar@{--}}
\newcommand{\dedge}{\ar@{-}}
\newcommand{\redge}{\ar@2{-}}
\newcommand{\eredgr}[1]{\mathcal{E}_{#1}}
\newcommand{\redgr}[1]{\mathcal{R}_{#1}}
\newcommand{\rgrem}[2]{\mathcal{R}_{rem_{#2}(#1)}}
\newcommand{\pcgr}{\mathcal{PC}}
\newcommand{\mergeEdges}{M}
\newcommand{\legalmset}{\omega}
\newcommand{\slegalmset}{\theta}
\newcommand{\flipm}{\mathrm{flip}}
\newcommand{\id}{\mathrm{id}}
\newcommand{\gredgr}{\mathcal{G}}
\newcommand{\range}{\mathrm{rng}}
\newcommand{\setlegalgredgr}{L}
\newcommand{\fiberonlabel}{C}
\newcommand{\bridge}{\mathrm{bridge}}
\newcommand{\dom}{\mathrm{dom}}
\newcommand{\odom}{\mathrm{odom}}
\renewcommand{\emptyset}{\varnothing}
\newcommand{\showfigs}[1]{#1}

}

\thesis{
\chapter{The Fibers and Range of Reduction Graphs}
}

\begin{abstract}
The biological process of gene assembly has been modeled based on
three types of string rewriting rules, called string pointer rules,
defined on so-called legal strings. It has been shown that reduction
graphs, graphs that are based on the notion of breakpoint graph in
the theory of sorting by reversal, for legal strings provide
valuable insights into the gene assembly process. We characterize
which legal strings obtain the same reduction graph (up to
isomorphism), and moreover we characterize which graphs are
(isomorphic to) reduction graphs.

More formally, let $\mathcal{R}$ be the function which assigns to
each legal string $u$ its reduction graph $\redgr{u}$. We
characterize the fiber $\mathcal{R}^{-1}(\redgr{u})$ (modulo graph
isomorphism) for each reduction graph $\redgr{u}$. In fact we show
that $\mathcal{R}^{-1}(\redgr{u})$ is the `orbit' of $u$ under two
types of string rewriting rules, which are in a way dual to two of
the three types of string pointer rules. We also characterize the
range of $\mathcal{R}$ in terms of easy-to-check conditions on
graphs.
\end{abstract}


\section{Introduction}
Ciliates form a large group of one-cellular organisms that are able
to transform one nucleus, called the micronucleus, into an
astonishing different one, called the macronucleus. This intricate
DNA transformation process is called \emph{gene assembly}. Each gene
in the micronucleus, called micronuclear gene, is transformed to a
gene in the macronucleus, called macronuclear gene. The string
pointer reduction system models gene assembly based on three types
of string rewriting rules, called string pointer rules, defined on
so-called legal strings~\cite{Ciliate_DB}. In this model, a
micronuclear gene is represented by a legal string $u$, while its
macronuclear gene (with its waste products) is represented by the
reduction graph of
$u$~\cite{Extended_paper,DBLP:conf/complife/BrijderHR05}. The
\emph{reduction graph} is based on the notion of breakpoint graph in
the theory of sorting by
reversal~\cite{DBLP:journals/jacm/HannenhalliP99,DBLP:conf/recomb/BergeronMS05,SetubalMeidanisBook}.
\symbolfootnote[0]{This research was supported by the Netherlands
Organization for Scientific Research (NWO) project 635.100.006
``VIEWS''.}

In this \paperchap we characterize which graphs are (isomorphic to)
reduction graphs (cf. Theorem~\ref{th_char_iso_redgr}). Obviously,
these graphs should have the `look and feel' of reduction graphs.
For instance, each vertex label should occur exactly four times, and
the second type of edges connect vertices of the same label. Once
these elementary properties are satisfied, reduction graphs are
characterized as having a connected pointer-component graph --- a
graph which represents the distribution of the vertex labels over
the connected components. The characterization corresponds to an
efficient algorithm. In this way we obtain a restriction on the form
of the macronuclear structures that can possibly occur. We also
provide a characterization that determines, given two legal strings,
whether or not they have the same reduction graph (cf.
Theorem~\ref{th_main_result}). This may allow one to determine which
micronuclear genes obtain the same macronuclear structure. It turns
out that two legal strings obtain the same reduction graph (up to
isomorphism) exactly when they can be transformed into each other by
two types of string rewriting rules, which surprisingly are in a
sense dual to the string positive rules and the string double rules
(two of the three types of string pointer rules).

The latter characterization has other uses as well. In a sense, the
reduction graph allows for a complete characterization of
applicability of \emph{string negative rules}, the other type of
string pointer rules, during the transformation
process~\cite{Extended_paper,StrategiesSnr/Brijder06,DBLP:conf/complife/BrijderHR06,DBLP:conf/complife/BrijderHR05}.
Moreover, it has been shown that the reduction graph does not retain
much information about the applicability of the other two types of
rules~\cite{StrategiesSnr/Brijder06}. Therefore, the legal strings
that obtain the same reduction graph are exactly the legal strings
that have similar characteristics concerning the string negative
rule.

To establish both main results, we augment the (abstract) reduction
graph with a set of \emph{merge-legal edges}. We will show that some
``valid'' sets of merge-legal edges for a reduction graph allows one
to ``go back'' to a legal string corresponding to this (abstract)
reduction graph. In this way the existence of such valid set
determines which graphs are (isomorphic to) reduction graphs. The
first main result shows that the existence of such valid set is
computationally easy to verify. Moreover, the set of all sets of
merge-legal edges can be transformed into each other by \emph{flip
operations}. These flip operations can be defined in terms of the
above mentioned dual string pointer rules on legal strings. This
will establish the other main result.

This \paperchap is organized as follows.
Section~\ref{sec_notation_terminology} fixes notation of basic
mathematical notions. In Section~\ref{sec_recall_sprs} we recall the
string pointer reduction system, in
Section~\ref{sec_recall_red_graph} we recall the reduction graph and
the pointer-component graph, and in
Section~\ref{sec_def_abstr_red_graph} we generalize the notion of
reduction graph and give an extension through merge-legal edges. In
Section~\ref{sec_ext_red_graph_to_legal_strings} we provide a
preliminary characterization that determines which graphs are
(isomorphic to) reduction graphs. In the next three sections, we
strengthen the result to allow for efficient algorithms: in
Section~\ref{sec_flip_edges} we define the flip operation on sets of
merge-legal edges, in Section~\ref{sec_flip_merging_split} we show
that the effect of flip operation corresponds to merging or
splitting of connected components, and in
Section~\ref{sec_conn_pc_graph} we prove the first main result, cf.
Theorem~\ref{th_char_iso_redgr}. In
Sections~\ref{sec_second_main_result1} and
\ref{sec_second_main_result2} we prove the second main result, cf.
Theorem~\ref{th_main_result}. We conclude this \paperchap with a
discussion.

\section{Mathematical Notation and Terminology}
\label{sec_notation_terminology} In this section we recall some
basic notions concerning functions, strings, and graphs. We do this
mainly to fix the basic notation and terminology.

The symmetric difference of sets $X$ and $Y$, $(X \backslash Y) \cup
(Y \backslash X)$, is denoted by $X \oplus Y$. The symmetric
difference of a finite family of sets $(X_i)_{i \in A}$ is denoted
by $\bigoplus_{i \in A} X_i$. The \emph{composition} of functions
$f: X \rightarrow Y$ and $g: Y \rightarrow Z$ is the function $g f:
X \rightarrow Z$ such that $(g f) (x) = g(f(x))$ for every $x \in
X$. The restriction of $f$ to a subset $A$ of $X$ is denoted by
$f|A$, . The range $f(X)$ of $f$ will be denoted by $\range(f)$. We
define for $y \in Y$, $f^{-1}(y) = \{ x \in X \mid f(x) = y \}$. If
$Y = X$, then $f$ is called \emph{self-inverse} if $f^2$ is the
identity function. We will use $\lambda$ to denote the empty string.


We now turn to graphs. A \emph{(undirected) graph} is a tuple $G =
(V,E)$, where $V$ is a finite set and $E \subseteq \{\{x,y\} \mid
x,y \in V\}$. The elements of $V$ are the \emph{vertices} of $G$ and
the elements of $E$ are the \emph{edges} of $G$. In this \paperchap
we allow $x = y$, and therefore edges can be of the form $\{x,x\} =
\{x\}$
--- an edge of this form should be seen as an edge connecting $x$ to
$x$, i.e., a `loop' for $x$. The \emph{restriction of $G$ to $E'
\subseteq E$}, denoted by $G|_{E'}$, is $(V,E')$. The \emph{order}
$|V|$ of $G$ is denoted by $o(G)$.

A \emph{multigraph} is a (undirected) graph $G = (V,E,\epsilon)$,
where parallel edges are possible. Therefore, $E$ is a finite set of
edges and $\epsilon: E \rightarrow \{\{x,y\} \mid x,y \in V\}$ is
the \emph{endpoint mapping}.

A \emph{coloured base} $B$ is a 4-tuple $(V,f,s,t)$ such that $V$ is
a finite set, $s,t \in V$, and $f: V \backslash \{s,t\} \rightarrow
\Gamma$ for some $\Gamma$. The elements of $V$, $\{\{x,y\} \mid x,y
\in V, x \not= y\}$, and $\Gamma$ are called \emph{vertices},
\emph{edges}, and \emph{vertex labels} for $B$, respectively.

A \emph{$n$-edge coloured graph}, $n \geq 1$, is a tuple $G =
(V,E_1,E_2,\cdots,E_n,f,s,t)$ where $B = (V,f,s,t)$ is a coloured
base and, for $i \in \{1,\ldots,n\}$, $E_i$ is a set of edges for
$B$. We also denote $G$ by $B(E_1,E_2,\cdots,E_n)$. We define
$\dom(G) = \range(f)$.

The previously defined notions and notation for graphs carry over to
multigraphs and $n$-edge coloured graphs. Isomorphisms between
graphs are defined in the usual way: they are considered
\emph{isomorphic} when they are equal modulo the identity of the
vertices. Thus, multigraphs $G = (V,E,\epsilon)$ and $G' =
(V',E,\epsilon')$ are \emph{isomorphic} if there is a bijection
$\alpha: V \rightarrow V'$ such that $\alpha \epsilon = \epsilon'$,
or more precisely, for $e \in E$, $\epsilon(e) = \{v_1,v_2\}$
implies $\epsilon'(e) = \{\alpha(v_1), \alpha(v_2)\}$. We assume the
reader is familiar with the notions of \emph{cycle} and
\emph{connected component} in a graph. A graph is called
\emph{connected} if it has exactly one connected component, and it
is called \emph{acyclic} when it does not contain cycles.

\section{String Pointer Reduction System} \label{sec_recall_sprs}
The string pointer reduction system is the model of gene assembly
that is used in this \paperchap. In this section we give a concise
description of this system, omitting examples and motivation. We
refer to \cite{GeneAssemblyBook} for an in-depth description of this
model including motivation and examples.

We fix $\kappa \geq 2$, and define the alphabet $\Delta =
\{2,3,\ldots,\kappa\}$. For $D \subseteq \Delta$, we define $\bar D
= \{ \bar a \mid a \in D \}$ and $\Pi = \Delta \cup \bar \Delta$.
The elements of $\Pi$ will be called \emph{pointers}. We use the
`bar operator' to move from $\Delta$ to $\bar \Delta$ and back from
$\bar \Delta$ to $\Delta$. Hence, for $p \in \Pi$, $\bar {\bar {p}}
= p$. For a string $u = x_1 x_2 \cdots x_n$ with $x_i \in \Pi$, the
\emph{inverse} of $u$ is the string $\bar u = \bar x_n \bar x_{n-1}
\cdots \bar x_1$. For $p \in \Pi$, we define $\pset{p} =
\begin{cases} p & \mbox{if } p \in \Delta \\ \bar{p} & \mbox{if }
p \in \bar{\Delta}
\end{cases}$, i.e., $\pset{p}$ is the `unbarred' variant of $p$. The
\emph{domain} of a string $v \in \Pi^*$ is $\dom(v) = \{ \pset{p}
\mid \mbox{$p$ occurs in $v$} \}$. A \emph{legal string} is a string
$u \in \Pi^*$ such that for each $p \in \Pi$ that occurs in $u$, $u$
contains exactly two occurrences from $\{p,\bar p\}$. For a pointer
$p$ and a legal string $u$, if both $p$ and $\bar p$ occur in $u$
then we say that both $p$ and $\bar p$ are \emph{positive} in $u$;
if on the other hand only $p$ or only $\bar p$ occurs in $u$, then
both $p$ and $\bar p$ are \emph{negative} in $u$.

Let $u = x_1 x_2 \cdots x_n$ be a legal string with $x_i \in \Pi$
for $1 \leq i \leq n$. For a pointer $p \in \Pi$ such that
$\{x_i,x_j\} \subseteq \{p,\bar p\}$ and $1 \leq i < j \leq n$, the
\emph{p-interval of $u$} is the substring $x_i x_{i+1} \cdots x_j$.
Two distinct pointers $p,q \in \Pi$ \emph{overlap} in $u$ if both
$\pset{q} \in dom(I_p)$ and $\pset{p} \in dom(I_q)$, where $I_p$
($I_q$, resp.) is the $p$-interval ($q$-interval, resp.) of $u$.

The string pointer reduction system consists of three types of
reduction rules, called \emph{string pointer rules}, operating on
legal strings. In this \paperchap we will not consider these rules
directly, but rather study the reduction graph (which is recalled in
the next section) that captures essential properties of the
rewriting system. For completeness we list the rules. For all $p,q
\in \Pi$ with $\pset{p} \not = \pset{q}$:
\begin{itemize}
\item
the \emph{string negative rule} for $p$ is defined by
$\textbf{snr}_{p}(u_1 p p u_2) = u_1 u_2$,
\item
the \emph{string positive rule} for $p$ is defined by
$\textbf{spr}_{p}(u_1 p u_2 \bar p u_3) = u_1 \bar u_2 u_3$,
\item
the \emph{string double rule} for $p,q$ is defined by
$\textbf{sdr}_{p,q}(u_1 p u_2 q u_3 p u_4 q u_5) = u_1 u_4 u_3 u_2
u_5$,
\end{itemize}
where $u_1,u_2,\ldots,u_5$ are arbitrary (possibly empty) strings
over $\Pi$.

We say that legal strings $u$ and $v$ are \emph{equivalent}, denoted
by $u \approx v$, if there is homomorphism $\varphi: \Pi^*
\rightarrow \Pi^*$ with $\varphi(p) \in \{p, \bar p\}$ and
$\varphi(\bar p) = \overline{\varphi(p)}$ for all $p \in \Pi$ such
that $\varphi(u) = v$.

\begin{Example}
Legal strings $2 \bar 2 3 3$ and $\bar 2 2 3 3$ are equivalent,
while $2 \bar 2 3 3$ are $2 \bar 2 \bar 3 3$ are not.
\end{Example}

Note that $\approx$ is an equivalence relation. Equivalent legal
strings are characterized by their `unbarred version' and their set
of positive pointers.

The \emph{domain} of a reduction rule $\rho$, denoted by
$\dom(\rho)$, is defined by $\dom(\textbf{snr}_p) =
\dom(\textbf{spr}_p) = \{\pset{p}\}$ and $\dom(\textbf{sdr}_{p,q}) =
\{\pset{p},\pset{q}\}$ for $p,q \in \Pi$. For a composition $\varphi
= \rho_n \ \cdots \ \rho_2 \ \rho_1$ of reduction rules $\rho_1,
\rho_2, \ldots, \rho_n$, the \emph{domain}, denoted by
$\dom(\varphi)$, is $\dom(\rho_1) \cup \dom(\rho_2) \cup \cdots \cup
\dom(\rho_n)$.

A composition $\varphi$ of reduction rules is called a
\emph{reduction}. Let $u$ be a legal string. We say that $\varphi$
is a \emph{reduction of $u$}, if $\varphi$ is a reduction and
$\varphi$ is applicable to (defined on) $u$. A \emph{successful
reduction $\varphi$ of $u$} is a reduction of $u$ such that
$\varphi(u) = \lambda$. We then also say that $\varphi$ is
\emph{successful for $u$}. For every legal string there exists a
successful reduction, which in general is not unique
\cite{GeneAssemblyBook}.

\section{Reduction Graph} \label{sec_recall_red_graph}
We now recall the definition of reduction graph. This definition is
equal to the one in \cite{StrategiesSnr/Brijder06}, and is in
slightly less general form compared to the one in
\cite{Extended_paper}. We refer to \cite{Extended_paper}, where it
was introduced, for a motivation and for more examples and results.
The notion of reduction graph uses the intuition from the notion of
breakpoint graph (or reality-and-desire diagram) known from another
branch of DNA processing theory called sorting by reversal, see e.g.
\cite{SetubalMeidanisBook} and \cite{PevznerBook}. From a biological
point of view, the reduction graph represents the macronuclear form
of a gene given its micronuclear form. This micronuclear form of the
gene is represented by a legal string, and therefore reduction
graphs are defined on legal strings.

\begin{Definition}
Let $u = p_1 p_2 \cdots p_n$ with $p_1,\ldots,p_n \in \Pi$ be a
legal string. The \emph{reduction graph of $u$}, denoted by
$\redgr{u}$, is a 2-edge coloured graph $(V,E_1,E_2,f,s,t)$, where
$$
V = \{\RGVertL{1},\RGVertL{2},\ldots,\RGVertL{n}\} \ \cup \
\{\RGVertR{1},\RGVertR{2},\ldots,\RGVertR{n}\} \ \cup \ \{s,t\},
$$
$$
E_{1} = \{e_0, e_1, \ldots, e_{n} \} \mbox{ with }  e_i = \{
\RGVertR{i},\RGVertL{i+1} \} \mbox{ for } 1 < i < n, e_0 =
\{s,\RGVertL{1}\}, e_n = \{ \RGVertR{n}, t \},
$$
\begin{eqnarray*}
E_{2} = & \{ \{\RGVertR{i},\RGVertL{j}\},
\{\RGVertL{i},\RGVertR{j}\} \ | \ i,j \in \{1,2,\ldots,n\}
\mbox{ with } i \not= j \mbox{ and } p_i = p_j \} \ \cup \ \\
& \{ \{\RGVertL{i},\RGVertL{j}\}, \{\RGVertR{i},\RGVertR{j}\} \ | \
i,j \in \{1,2,\ldots,n\} \mbox{ and } p_i = \bar{p}_j \}, \mbox{
and}
\end{eqnarray*}
$$
\mbox{$f(\RGVertL{i}) = f(\RGVertR{i}) = \pset{p_i}$ for $1 \leq i
\leq n$.}
$$
\mbox{  }
\end{Definition}
The edges of $E_1$ are called the \emph{reality edges}, and the
edges of $E_2$ are called the \emph{desire edges}. Notice that for
each $p \in \dom(u)$, the reduction graph of $u$ has exactly two
desire edges containing vertices labelled by $p$. It follows from
the construction of the reduction graph that, given legal strings
$u$ and $v$, $u \approx v$ implies that $\redgr{u} \approx
\redgr{v}$.

In depictions of reduction graphs, we will represent the vertices
(except for $s$ and $t$) by their labels, because the exact identity
of the vertices is not essential for the problems considered in this
\paperchap. We will also depict reality edges as `double edges' to
distinguish them from the desire edges.

\showfigs{
\begin{figure}
\scalebox{0.65}{
$$
\xymatrix@=10pt{
\\
\\
\\
\scalebox{1.3}{s} \redge[r] &
2 \ar@{-}@/^4pc/[rrrrrrrrrrrrrrrrr] & 2 \redge[r]
\ar@{-}@/^3.5pc/[rrrrrrrrrrrrrrr] &
7 \ar@{-}@/_1pc/[rrrr] & 7 \redge[r] \ar@{-}@/_1pc/[rrrr] &
4 \ar@{-}@/^2.5pc/[rrrrrrrrrr] & 4 \redge[r]
\ar@{-}@/^2.5pc/[rrrrrrrrrr] &
7 & 7 \redge[r] &
3 \ar@{-}@/^1.7pc/[rrrrr] & 3 \redge[r] \ar@{-}@/^1.3pc/[rrr] &
5 \ar@{-}@/_2.5pc/[rrrrrrrrrrr] & 5 \redge[r]
\ar@{-}@/_2pc/[rrrrrrrrr] &
3 & 3 \redge[r] &
4 & 4 \redge[r] &
2 & 2 \redge[r] &
6 \ar@{-}@/^1.8pc/[rrrrr] & 6 \redge[r] \ar@{-}@/^1.2pc/[rrr] &
5 & 5 \redge[r] &
6 & 6 \redge[r] &
\scalebox{1.3}{t}
\\
\\
\\
}
$$
} \caption{The reduction graph $\redgr{u}$ of $u$ in
Example~\ref{ex_abstr_red_graph1}.} \label{ex_abstr_red_graph1_0}
\end{figure}
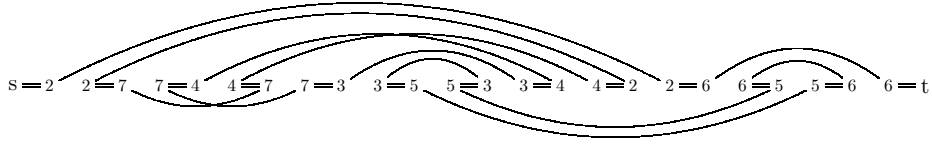
}

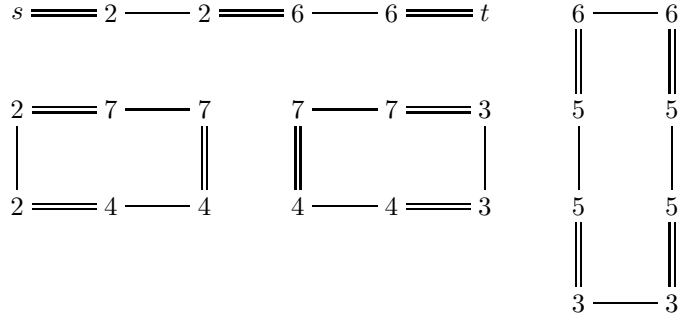
\begin{figure}
$$ \xymatrix{
s \redge[r] & 2 \dedge[r]  & 2 \redge[r] & 6 \dedge[r]  & 6
\redge[r] & t & 6 \dedge[r] \redge[d] & 6 \redge[d]\\
2 \redge[r] \dedge[d] & 7 \dedge[r] & 7 \redge[d] & 7 \redge[d] \dedge[r] & 7 \redge[r] & 3 \dedge[d] & 5 \dedge[d] & 5 \dedge[d] \\
2 \redge[r]           & 4 \dedge[r] & 4           & 4 \dedge[r] & 4 \redge[r] & 3 & 5 \redge[d] & 5 \redge[d] \\
& & & & & & 3 \dedge[r] & 3 } $$ \caption{The reduction graph of
Figure~\ref{ex_abstr_red_graph1_0} obtained by rearranging the
vertices.} \label{ex_abstr_red_graph1_1}
\end{figure}

\begin{Example} \label{ex_abstr_red_graph1}
The reduction graph of $u = 2 \bar 7 4 7 3 5 3 \bar 4 2 6 5 6$ is
depicted in Figure~\ref{ex_abstr_red_graph1_0}. Note how positive
pointers are connected by crossing desire edges, while those for
negative pointers are parallel. By rearranging the vertices we can
depict the graph as shown in Figure~\ref{ex_abstr_red_graph1_1}.
\end{Example}

Reality edges follow the linear order of the legal string, whereas
desire edges connect positions in the string that will be joined
when performing reduction rules, see \cite{Extended_paper}.

We now recall the definition of pointer-component graph of a legal
string, introduced in \cite{StrategiesSnr/Brijder06}. Surprisingly
however, this graph has different uses in this \paperchap compared
to its original uses in \cite{StrategiesSnr/Brijder06}, where it is
used to characterize which string negative rules are used in
successful reductions of the legal string.
\begin{Definition}
Let $u$ be a legal string. The \emph{pointer-component graph of $u$
(or of $\redgr{u}$)}, denoted by $\pcgr_u$, is a multigraph $(\zeta,
E, \epsilon)$, where $\zeta$ is the set of connected components of
$\redgr{u}$, $E = \dom(u)$ and $\epsilon$ is, for $e \in E$, defined
by $\epsilon(e) = \{C \in \zeta \mid C$ $\mbox{contains}$
$\mbox{vertices}$ $\mbox{labelled by } e\}$.
\end{Definition}

\begin{figure}
$$ \xymatrix{
C_1 \ar@(u,l)@{-}[]_5 \ar@{-}[r]^6 & R \ar@{-}[d]^2 \\
C_2 \ar@{-}[u]^3 \ar@/^/@{-}[r]^4 & C_3 \ar@/^/@{-}[l]^7
} $$ \caption{The pointer-component graph of the reduction graph
from Figure~\ref{ex_abstr_red_graph1_1}.}
\label{ex_abstr_red_graph1_2}
\end{figure}
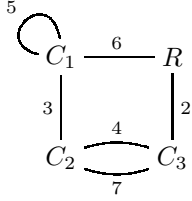

\begin{Example}
The pointer-component graph of the reduction graph from
Figure~\ref{ex_abstr_red_graph1_1} is shown in
Figure~\ref{ex_abstr_red_graph1_2}.
\end{Example}

\section{Abstract Reduction Graphs and Extensions}
\label{sec_def_abstr_red_graph} In this section we generalize the
notion of reduction graph as a starting point to consider which
graphs are (isomorphic to) reduction graphs. Moreover, we extend the
reduction graphs by a set of edges, called \emph{merge edges}, such
that, along with the reality edges, the linear structure of the
legal string is preserved in the graph.

We will now define a set of edges for a given coloured base which
has features in common with desire edges of a reduction graph.
\begin{Definition}
Let $B = (V,f,s,t)$ be a coloured base. We say that a set of edges
$E$ for $B$ is \emph{desirable} if
\begin{enumerate}
\item
for all $\{v_1, v_2\} \in E$, $f(v_1) = f(v_2)$,
\item
for each $v \in V \backslash \{s,t\}$ there is exactly one $e \in E$
such that $v \in e$.
\end{enumerate}
\end{Definition}

We now generalize the concept of reduction graph.
\begin{Definition}
A 2-edge coloured graph $B(E_1,E_2)$ with $B = (V,f,s,t)$ is called
an \emph{abstract reduction graph} if
\begin{enumerate}
\item
$\range(f) \subseteq \Delta$, and for each $p \in \range(f)$,
$|f^{-1}(p)| = 4$,
\item
for each $v \in V$ there is exactly one $e \in E_1$ such that $v \in
e$,
\item
$E_2$ is desirable for $B$.
\end{enumerate}
\end{Definition}
The set of all abstract reduction graphs is denoted by $\gredgr$.

Clearly, if $G \approx \redgr{u}$ for some $u$, then $G \in
\gredgr$. Therefore, for abstract reduction graphs $G = B(E_1,E_2)$,
the edges in $E_1$ are called \emph{reality edges} and the edges in
$E_2$ are called \emph{desire edges}. For graphical depictions of
abstract reduction graphs we will use the same conventions as we
have for reduction graphs. Thus, edges in $E_1$ will be depicted as
``double edges'', vertices are represented by their label, etc.

\begin{figure}
$$ \xymatrix{
2 \redge[d] \dedge[r] & 2 \redge[d] & 5 \redge[d] \dedge[r] & 5 \redge[r] & 9 \dedge[d] & 8 \redge[d] \dedge[r] & 8 \redge[d] & & s \redge[d] \\
5           \dedge[r] & 5           & 4           \dedge[r] & 4 \redge[r] & 9           & 7 \dedge[r]           & 7           & & 9 \dedge[d] \\
2 \redge[d] \dedge[r] & 2 \redge[d] & 3 \redge[d] \dedge[r] & 3 \redge[r] & 8 \dedge[r] & 8 \redge[r]           & 3 \dedge[r] & 3 \redge[d] & 9 \redge[d] \\
4           \dedge[r] & 4           & 7           \dedge[r] & 7 \redge[r] & 6 \dedge[r] & 6 \redge[r]           & 6 \dedge[r] & 6 & t           \\
} $$ \caption{An abstract reduction graph.}
\label{ex_abstr_red_graph2_1}
\end{figure}

\begin{Example}
The 2-edge coloured graph in Figure~\ref{ex_abstr_red_graph2_1} is
an abstract reduction graph.
\end{Example}

Note that conditions (1) and (3) in the previous definition imply
that for each $p \in \range(f)$, there is a partition $\{e_1,e_2\}$
of $f^{-1}(p)$, denoted by $\fiberonlabel_{G,p}$ or
$\fiberonlabel_{p}$ when $G$ is clear from the context, such that
$e_1, e_2 \in E_2$.

We now introduce an extension to reduction graphs such that the
`generic' linear order of the vertices $s, \RGVertL{1}, \RGVertR{1},
\ldots, \RGVertL{n}, \RGVertR{n}, t$ is retained, even when we
consider the graphs up to isomorphism.
\begin{Definition}
Let $u$ be a legal string. The \emph{extended reduction graph of
$u$}, denoted by $\eredgr{u}$, is a 3-edge coloured graph
$B(E_1,E_2,E_3)$, where $\redgr{u} = B(E_1,E_2)$ and $E_3 = \{
\{\RGVertL{i},\RGVertR{i}\} \mid 1 \leq i \leq n \}$ with $n = |u|$.
\end{Definition}
The edges in $E_3$ are called the \emph{merge edges of $u$}, denoted
by $\mergeEdges_u$. In this way, the reality edges and the merge
edges form a unique path which passes through the vertices in the
generic linear order. This is illustrated in the next example. In
figures merge edges will be depicted by ``dashed edges''.

\showfigs{
\begin{figure}
\scalebox{0.48}{
$$
\xymatrix@=15pt{
\\
\\
\scalebox{1.5}{s} \redge[r] &
\scalebox{1.3}{2} \ar@{-}@/^4.3pc/[rrrrrrrrrrrrrrrrr] \medge[r] &
\scalebox{1.3}{2} \ar@{-}@/^3.8pc/[rrrrrrrrrrrrrrr] \redge[r]  &
\scalebox{1.3}{7} \ar@{-}@/_1.3pc/[rrrr] \medge[r] &
\scalebox{1.3}{7} \redge[r] \ar@{-}@/_1.3pc/[rrrr] &
\scalebox{1.3}{4} \ar@{-}@/^2.8pc/[rrrrrrrrrr] \medge[r] &
\scalebox{1.3}{4} \redge[r] \ar@{-}@/^2.8pc/[rrrrrrrrrr] &
\scalebox{1.3}{7} \medge[r] & \scalebox{1.3}{7} \redge[r] &
\scalebox{1.3}{3} \ar@{-}@/^2pc/[rrrrr] \medge[r] &
\scalebox{1.3}{3} \redge[r] \ar@{-}@/^1.6pc/[rrr] &
\scalebox{1.3}{5} \ar@{-}@/_2.8pc/[rrrrrrrrrrr] \medge[r] &
\scalebox{1.3}{5} \redge[r] \ar@{-}@/_2.3pc/[rrrrrrrrr] &
\scalebox{1.3}{3} \medge[r] & \scalebox{1.3}{3} \redge[r] &
\scalebox{1.3}{4} \medge[r] & \scalebox{1.3}{4} \redge[r] &
\scalebox{1.3}{2} \medge[r] & \scalebox{1.3}{2} \redge[r] &
\scalebox{1.3}{6} \ar@{-}@/^2pc/[rrrrr] \medge[r] &
\scalebox{1.3}{6} \redge[r] \ar@{-}@/^1.6pc/[rrr] &
\scalebox{1.3}{5} \medge[r] & \scalebox{1.3}{5} \redge[r] &
\scalebox{1.3}{6} \medge[r] & \scalebox{1.3}{6} \redge[r] &
\scalebox{1.5}{t}
\\
\\
}
$$
} \caption{The extended reduction graph $\eredgr{u}$ of $u$ given in
Example~\ref{ex_abstr_red_graph1}.} \label{ex_abstr_red_graph1_3}
\end{figure}
}

\begin{Example}
The extended reduction graph $\eredgr{u}$ of $u$ given in
Example~\ref{ex_abstr_red_graph1} is shown in
Figure~\ref{ex_abstr_red_graph1_3}, cf.
Figure~\ref{ex_abstr_red_graph1_0}.
\end{Example}

\begin{Remark}
The notion of merge edges for (extended) reduction graphs is more
closely related to the notion of reality edges for breakpoint graphs
in the theory of sorting-by-reversal compared to the notion of
reality edges for (extended) reduction graphs. Thus in a way it
would be more natural to call the merge edges reality edges for
(extended) reduction graphs, and the other way around. However, to
avoid confusion with earlier work, we do not change this
terminology.
\end{Remark}

We now generalize this extension of reduction graphs to abstract
reduction graphs.
\begin{Definition} \label{def_merge_legal}
Let $G = B(E_1,E_2) \in \gredgr$, and let $E$ be a set of edges for
$B$. We say that $E$ is \emph{merge-legal} for $G$ if $E$ is
desirable for $B$, and $E_2 \cap E = \emptyset$. We denote the set
$\{E \mid E \mbox{ merge-legal for } G\}$ by $\legalmset_G$. The set
of all $E \in \legalmset_G$ where $B(E_1,E)$ is connected is denoted
by $\slegalmset_G$.
\end{Definition}
For legal string $u$, we also denote $\legalmset_{\redgr{u}}$ and
$\slegalmset_{\redgr{u}}$ by $\legalmset_{u}$ and $\slegalmset_{u}$,
respectively.

Notice that $\mergeEdges_u \in \slegalmset_u \subseteq
\legalmset_{u}$. Therefore, merge-legal edges will also be depicted
by ``dashed edges''.

\begin{figure}
$$ \xymatrix{
s \redge[r] & 2 \dedge[r] & 2 \redge[r] & 3 \dedge[r] & 3 \redge[r]
& t \\
& & 2 \dedge[d] \redge[r] & 3 \dedge[d] \\
& & 2 \redge[r] & 3 \\
} $$ \caption{An abstract reduction graph.}
\label{ex_abstr_red_graph3_1}
\end{figure}
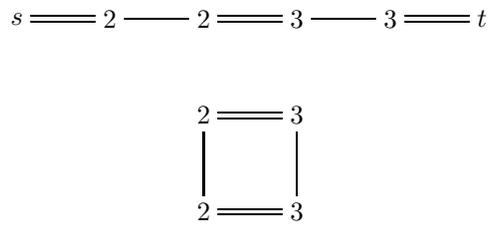

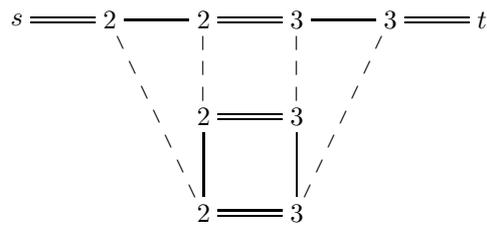
\begin{figure}
$$ \xymatrix{
s \redge[r] & 2 \dedge[r] \medge[rdd] & 2 \medge[d] \redge[r] & 3
\dedge[r] \medge[d] & 3 \redge[r] \medge[ddl]
& t \\
& & 2 \dedge[d] \redge[r] & 3 \dedge[d] \\
& & 2 \redge[r] & 3 \\
} $$ \caption{The abstract reduction graph of
Figure~\ref{ex_abstr_red_graph3_1} with a set of merge-legal edges.}
\label{ex_abstr_red_graph3_2}
\end{figure}

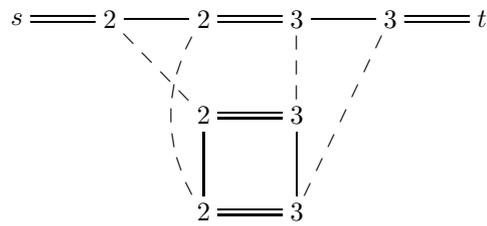
\begin{figure}
$$ \xymatrix{
s \redge[r] & 2 \dedge[r] \medge[rd] & 2 \ar@{--}@/_1pc/[dd]
\redge[r] & 3 \dedge[r] \medge[d] & 3 \redge[r] \medge[ddl]
& t \\
& & 2 \dedge[d] \redge[r] & 3 \dedge[d] \\
& & 2 \redge[r] & 3 \\
} $$ \caption{The abstract reduction graph of
Figure~\ref{ex_abstr_red_graph3_1} with another set of merge-legal
edges.} \label{ex_abstr_red_graph3_3}
\end{figure}

\begin{Example}
Let us consider the abstract reduction graph $G = B(E_1,E_2)$ of
Figure~\ref{ex_abstr_red_graph3_1}. This graph is again depicted in
Figure~\ref{ex_abstr_red_graph3_2} including a merge-legal set $E$
for $G$. In this way Figure~\ref{ex_abstr_red_graph3_2} depicts the
3-edge coloured graph $B(E_1,E_2,E)$. Notice that $E \not\in
\slegalmset_G$. In Figure~\ref{ex_abstr_red_graph3_3}, the abstract
reduction graph is depicted with a merge-legal set in
$\slegalmset_G$.
\end{Example}

We now define a natural abstraction of the notion of extended
reduction graph.
\begin{Definition}
Let $G = B(E_1,E_2) \in \gredgr$ and $E \in \slegalmset_{G}$. Then
$G' = B(E_1,E_2,E)$ is called a \emph{extended abstract reduction
graph}.
\end{Definition}
For each legal string $u$, $\eredgr{u}$ is an extended abstract
reduction graph, since $\mergeEdges_u \in \slegalmset_u$. Therefore,
the edges in $E$ (in the previous definition) are called the
\emph{merge edges (of $G'$)}. Since $E \in \slegalmset_{G}$,
$B(E_1,E)$ has the following form:
$$
\xymatrix{
s \redge[r] & \pset{p}_1 \medge[r] & \pset{p}_1 \redge[r] &
\pset{p}_2 \medge[r] & \pset{p}_2 \redge[r] & \cdots \redge[r] &
\pset{p}_n \medge[r] & \pset{p}_n \redge[r] & t
}
$$
Thus the property that reality and merge edges in an extended
reduction graph induce a unique path from $s$ to $t$ that
alternatingly passes through reality edges and merge edges is
retained for extended abstract reduction graphs $G$ in general.

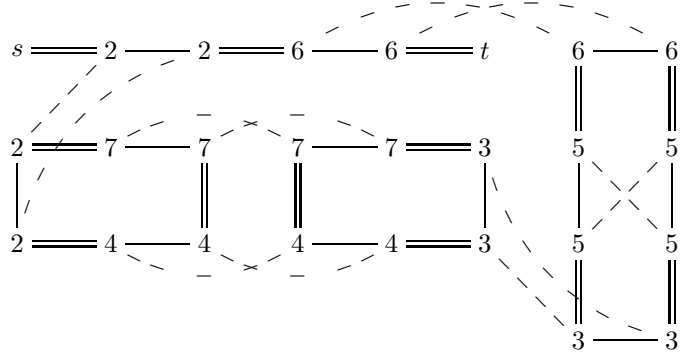
\begin{figure}
$$ \xymatrix{
\\
\\
s \redge[r] & 2 \dedge[r] \medge[ld] & 2 \redge[r]
\medge@/_1.3pc/[lldd] & 6 \dedge[r] \medge@/^1.5pc/[rrr]  & 6
\medge@/^1.5pc/[rrr]
\redge[r] & t & 6 \dedge[r] \redge[d] & 6 \redge[d]\\
2 \redge[r] \dedge[d] & 7 \medge@/^1pc/[rr] \dedge[r] & 7 \redge[d] \medge@/^1pc/[rr] & 7 \redge[d] \dedge[r] & 7 \redge[r] & 3 \dedge[d] & 5 \dedge[d] \medge[rd] & 5 \dedge[d] \medge[ld] \\
2 \redge[r]           & 4 \medge@/_1pc/[rr] \dedge[r] & 4 \medge@/_1pc/[rr]  & 4 \dedge[r] & 4 \redge[r] & 3 & 5 \redge[d] & 5 \redge[d] \\
& & & & & & 3 \dedge[r] \medge[lu] & 3 \medge@/^1.3pc/[lluu] }
$$ \caption{A extended abstract reduction graph obtained by
augmenting the reduction graph of Figure~\ref{ex_abstr_red_graph1_1}
with merge edges.} \label{ex_abstr_red_graph1_4}
\end{figure}

\begin{Example}
If we consider the reduction graph $\redgr{u} = B(E_1,E_2)$ of
Example~\ref{ex_abstr_red_graph1} shown in
Figure~\ref{ex_abstr_red_graph1_1}, then, of course,
$B(E_1,E_2,\mergeEdges_u) = \eredgr{u}$ shown in
Figure~\ref{ex_abstr_red_graph1_3} is a extended abstract reduction
graph. In Figure \ref{ex_abstr_red_graph1_4} another extended
reduction graph is shown -- it is $\redgr{u}$ augmented with a set
of merge edges $E$ in $\slegalmset_u$. It is easy to see that indeed
$E \in \slegalmset_u$: simply notice that the path from $s$ to $t$
induced by the reality and merge edges will go through every vertex
of the graph.
\end{Example}

\section{Back to Legal Strings}
\label{sec_ext_red_graph_to_legal_strings} In this section we show
that for extended abstract reduction graphs $G$ we can `go back' in
the sense that there are legal strings $u$ such that $G$ is
isomorphic to $\eredgr{u}$. Moreover we show how to obtain the set
$\setlegalgredgr_G$ of all legal strings that corresponds to $G$. We
will show that the legal strings in $\setlegalgredgr_G$ are
equivalent, and thus that extended reduction graphs retain all
essential information of the legal strings.

As extended abstract reduction graphs have a natural linear order of
the vertices given by their reality edges and merge edges, we can
infer whether or not desire edges `cross' or not. Thereby providing
a way to define negative and positive pointers for extended abstract
reduction graphs.
\begin{Definition} \label{def_neg_pos_ex_abstr_red_graph}
Let $G = B(E_1,E_2,E_3)$ be an extended abstract reduction graph,
let $G' = B(E_1,E_2)$, and let $\pi =
(s,v_1,v'_1,\cdots,v_n,v'_{n},t)$ be the path from $s$ to $t$ in
$B(E_1,E_3)$. We say that $p \in \dom(G)$ is \emph{negative in $G$}
iff $\fiberonlabel_{G',p} = \{\{v_{i}, v'_{j}\}, \{v'_{i},
v_{j}\}\}$ for some $i,j \in \{1, \ldots, n\}$ with $i \not= j$.
Also, we say that $p \in \dom(G)$ is \emph{positive in $G$} if $p$
is not negative in $G$.
\end{Definition}
Clearly, $p \in \dom(G)$ is positive in $G$ iff
$\fiberonlabel_{G',p} = \{\{v_{i}, v_{j}\}, \{v'_{i}, v'_{j}\}\}$
for some $i,j \in \{1, \ldots, n\}$ with $i \not= j$. It is easy to
see that $p$ is negative in legal string $u$ iff $p$ is negative in
$\eredgr{u}$.

The next definition defines a set of legal strings for each extended
abstract reduction graph.
\begin{Definition} \label{def_legalization}
Let $G = B(E_1,E_2,E_3)$ be an extended abstract reduction graph,
let $G' = B(E_1,E_2)$, and let $H = B(E_1,E_3)$ be as follows:
$$
\xymatrix{
s \redge[r] & \pset{p}_1 \medge[r] & \pset{p}_1 \redge[r] &
\pset{p}_2 \medge[r] & \pset{p}_2 \redge[r] & \cdots \redge[r] &
\pset{p}_n \medge[r] & \pset{p}_n \redge[r] & t
}
$$
The \emph{legalization} of $G$, denoted by $\setlegalgredgr_G$, is
the set of legal strings $u = p_1 p_2 \cdots p_{n}$ with $p_i \in
\{\pset{p}_i, \overline{\pset{p}}_i\}$ and $p_i$ is negative in $u$
iff $p_i$ is negative in $G$.
\end{Definition}

\showfigs{
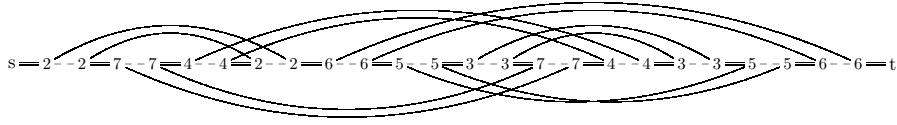
\begin{figure}
\scalebox{0.48}{
$$
\xymatrix@=15pt{
\\
\\
\\
\scalebox{1.5}{s} \redge[r] &
\scalebox{1.3}{2} \ar@{-}@/^2.5pc/[rrrrrrr] \medge[r] &
\scalebox{1.3}{2} \ar@{-}@/^2pc/[rrrrr] \redge[r]  &
\scalebox{1.3}{7} \ar@{-}@/_3.5pc/[rrrrrrrrrrrrr] \medge[r] &
\scalebox{1.3}{7} \redge[r] \ar@{-}@/_3pc/[rrrrrrrrrrr] &
\scalebox{1.3}{4} \ar@{-}@/^3.5pc/[rrrrrrrrrrrrr] \medge[r] &
\scalebox{1.3}{4} \redge[r] \ar@{-}@/^3pc/[rrrrrrrrrrr] &
\scalebox{1.3}{2} \medge[r] & \scalebox{1.3}{2} \redge[r] &
\scalebox{1.3}{6} \ar@{-}@/^4pc/[rrrrrrrrrrrrrrr] \medge[r] &
\scalebox{1.3}{6} \redge[r] \ar@{-}@/^3.5pc/[rrrrrrrrrrrrr] &
\scalebox{1.3}{5} \ar@{-}@/_2.5pc/[rrrrrrrrrr] \medge[r] &
\scalebox{1.3}{5} \redge[r] \ar@{-}@/_2.5pc/[rrrrrrrrrr] &
\scalebox{1.3}{3} \ar@{-}@/^2.5pc/[rrrrrrr] \medge[r] &
\scalebox{1.3}{3} \redge[r] \ar@{-}@/^2pc/[rrrrr] &
\scalebox{1.3}{7} \medge[r] & \scalebox{1.3}{7} \redge[r] &
\scalebox{1.3}{4} \medge[r] & \scalebox{1.3}{4} \redge[r] &
\scalebox{1.3}{3} \medge[r] & \scalebox{1.3}{3} \redge[r] &
\scalebox{1.3}{5} \medge[r] & \scalebox{1.3}{5} \redge[r] &
\scalebox{1.3}{6} \medge[r] & \scalebox{1.3}{6} \redge[r] &
\scalebox{1.5}{t}
\\
\\
\\
}
$$
} \caption{The extended abstract reduction graph $G$ given in
Example~\ref{ex_ext_abstr_red_graph}.} \label{ex_abstr_red_graph1_5}
\end{figure}
}

\begin{Example} \label{ex_ext_abstr_red_graph}
Let us consider the extended abstract reduction graph $G$ of
Figure~\ref{ex_abstr_red_graph1_4}. By rearranging the vertices we
obtain Figure~\ref{ex_abstr_red_graph1_5}. From this figure it is
clear that $v = 2 7 4 8 8 2 6 \bar 5 3 7 4 3 5 6 \in
\setlegalgredgr_G$.
\end{Example}

It is easy to see that, for a legal string $u$, we have $u \in
\setlegalgredgr_{\eredgr{u}}$.

Note that $\setlegalgredgr_G$, for extended abstract reduction graph
$G$, is an non-empty equivalence class w.r.t. to the $\approx$
relation (for legal strings). Since the definition of
$\setlegalgredgr_G$ does not depend on the exact identity of the
vertices of $G$, we have, for extended abstract reduction graphs $G$
and $G'$, $G \approx G'$ implies $\setlegalgredgr_{G} =
\setlegalgredgr_{G'}$.

\begin{Theorem} \label{th_legal_unique_ext}
\begin{enumerate}
\item Let $G$ and $G'$ be extended abstract reduction graphs. Then $G
\approx G'$ iff $\setlegalgredgr_{G} = \setlegalgredgr_{G'}$.
\item Let $u$ and $v$ be legal strings. Then $u \approx v$ iff $\eredgr{u}
\approx \eredgr{v}$.
\end{enumerate}
\end{Theorem}
\begin{Proof}
We first consider statement 1. We have already established the
forward implication. We now prove the reverse implication. Let $G =
B(E_1,E_2,E_3)$, $G' = B'(E'_1,E'_2,E'_3)$, and $\setlegalgredgr_{G}
= \setlegalgredgr_{G'}$. By the definition of legalization,
$B(E_1,E_3) \approx B'(E'_1,E'_3)$ and $p$ is negative in $G$ iff
$p$ is negative in $G'$ for $p \in dom(G) = dom(G')$. Therefore, $G
\approx G'$.

We now consider statement 2. We have $u \approx v$ iff $u,v \in
\setlegalgredgr_{\eredgr{u}} = \setlegalgredgr_{\eredgr{v}}$ (since
legalizations are equivalence classes of legal strings w.r.t
$\approx$) iff $\eredgr{u} \approx \eredgr{v}$ (by the first
statement).
\end{Proof}

Let $G$ be an extended abstract reduction graph, and take $u \in
\setlegalgredgr_{G}$ (such a $u$ exists since $\setlegalgredgr_{G}$
is nonempty). Since $u \in \setlegalgredgr_{\eredgr{u}}$ and
legalizations are equivalence classes, we have
$\setlegalgredgr_{\eredgr{u}} = \setlegalgredgr_{G}$ and therefore
$G \approx \eredgr{u}$. Thus every extended abstract reduction graph
$G$ is isomorphic to an extended reduction graph. In fact, it is
isomorphic to precisely those extended reduction graphs $\eredgr{u}$
with $u \in \setlegalgredgr_G$. Therefore, this $u$ is unique up to
equivalence.

\begin{Corollary} \label{cor_iso_redgraph_merge}
Let $u$ and $v$ be legal strings. If $\redgr{u} \approx \redgr{v}$,
then there is a $E \in \slegalmset_u$ such that $\eredgr{v} \approx
B(E_1,E_2,E)$ with $\redgr{u} = B(E_1,E_2)$.
\end{Corollary}
\begin{Proof}
Since $\redgr{u} \approx \redgr{v}$, there is an set of edges $E$
for $\redgr{u}$ such that $\eredgr{v} \approx B(E_1,E_2,E)$. Since
$\mergeEdges_v \in \slegalmset_v$, we have $E \in \slegalmset_u$.
\end{Proof}

We end this section with a graph theoretical characterization of
reduction graphs.
\begin{Theorem} \label{th_iso_gredgr_redgr1}
Let $G$ be a $2$-edge coloured graph. Then $G$ is isomorphic to a
reduction graph iff $G \in \gredgr$ and $\slegalmset_G \not=
\emptyset$.
\end{Theorem}
\begin{Proof}
Let $G \approx \redgr{u}$ for some legal string $u$. Then clearly,
$G \in \gredgr$. Also, $\mergeEdges_u \in \slegalmset_u$ and hence
$\slegalmset_u \not= \emptyset$. Therefore, $\slegalmset_G \not=
\emptyset$.

Let $E \in \slegalmset_G$. Then $G' = B(E_1,E_2,E)$ is an extended
abstract reduction graph with $G = B(E_1,E_2)$. By the paragraph
below Theorem~\ref{th_legal_unique_ext}, $G' \approx \eredgr{u}$ for
some legal string $u$ (take $u \in \setlegalgredgr_{G'}$). Hence, $G
\approx \redgr{u}$.
\end{Proof}

\section{Flip Edges} \label{sec_flip_edges}
In this section and the next two we provide characterizations of the
statement $\slegalmset_G \not= \emptyset$. This allows, using
Theorem~\ref{th_iso_gredgr_redgr1}, for a characterization that
corresponds to an efficient algorithm that determines whether or not
a given $G \in \gredgr$ is isomorphic to a reduction graph.
Moreover, it allows for an efficient algorithm that determines a
legal string $u$ for which $G \approx \redgr{u}$.

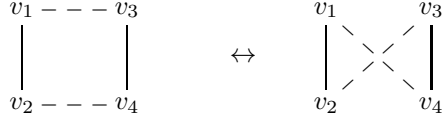
\begin{figure}
\begin{center}
\setlength{\unitlength}{1cm} 
\begin{picture}(7,2)(-0.5,-0.5)
\put(0,1) {$\xymatrix{
v_1 \medge[r] \dedge[d] & v_3 \dedge[d] \\
v_2 \medge[r] & v_4 }$}
\put(3,0.4) {$\leftrightarrow$}
\put(4,1) {$\xymatrix{
v_1 \medge[rd] \dedge[d] & v_3 \dedge[d] \\
v_2 \medge[ru] & v_4 }$} \end{picture}
\end{center}
\caption{Flip operation for $p$. All vertices are labelled by $p$}
\label{ex1_flip_merge}
\end{figure}

Let $G \in \gredgr$. Then a merge-legal set for $G$ is easily
obtained. For each $p \in \dom(G)$ with $\fiberonlabel_p = \{ \{v_1,
v_2\}, \{v_3,v_4\} \}$, a merge-legal set for $G$ must have either
the edges $\{v_1,v_3\}$ and $\{v_2,v_4\}$ or the edges $\{v_1,v_4\}$
and $\{v_2,v_3\}$, see both sides in Figure~\ref{ex1_flip_merge}. By
assigning such edges for each $p \in \dom(G)$ we obtain a
merge-legal set for $G$. Thus, $\legalmset_G \not= \emptyset$ for
each $G \in \gredgr$. Note that in particular, if $\dom(G) =
\emptyset$, then $\legalmset_G = \{\emptyset\}$. However,
$\slegalmset_G$ can be empty as the next example will illustrate.

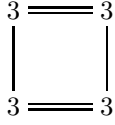
\begin{figure}
$$ \xymatrix{
s \redge[r] & 2 \dedge[r] & 2 \redge[r] & 2 \dedge[r] & 2 \redge[r]
& t \\
& & 3 \dedge[d] \redge[r] & 3 \dedge[d] \\
& & 3 \redge[r] & 3 \\
} $$ \caption{An abstract reduction graph $G$ for which
$\slegalmset_G = \emptyset$.} \label{ex_abstr_red_graph4_1}
\end{figure}
\begin{Example}
It is easy to see that the abstract reduction graph $G$ of
Figure~\ref{ex_abstr_red_graph4_1} does not have a merge-legal set
in $\slegalmset_G$.
\end{Example}

We now formally define a type of operation that in
Figure~\ref{ex1_flip_merge} transforms the situation on the
left-hand side to the situation on the right-hand side, and the
other way around. Informally speaking it ``flips'' edges of
merge-legal sets.

\begin{Definition}
Let $G = B(E_1,E_2) \in \gredgr$, let $f$ be the vertex labeling
function of $G$, and let $p \in \dom(G)$. The \emph{flip operation
for $p$ (w.r.t. $G$)}, denoted by $\flipm_{G,p}$, is the function
$\legalmset_G \rightarrow \legalmset_G$ defined by:
$$
\flipm_{G,p}(E) = \{ \{v_1,v_2\} \in E \mid f(v_1) \not= p \not=
f(v_2) \} \cup \{e_1, e_2\},
$$
where $e_1$ and $e_2$ are the two edges with vertices labelled by
$p$ such that $e_1, e_2 \not\in E_2 \cup E$.
\end{Definition}
When $G$ is clear from the context, we also denote $\flipm_{G,p}$ by
$\flipm_p$.

Since by Figure~\ref{ex1_flip_merge}, there are exactly two edges
$e_1$ and $e_2$ with vertices labelled by $p$ that are not parallel
to both the edges in $E_2 \cup E$, $\flipm_p$ is well defined. It is
now easy to see that indeed $\flipm_p(E) \in \legalmset_G$ for $E
\in \legalmset_G$.

\begin{Example}
Let $G$ be the abstract reduction graph of
Figure~\ref{ex_abstr_red_graph3_1}. If we apply $\flipm_{G,2}$ to
the set of merge-legal edges depicted in
Figure~\ref{ex_abstr_red_graph3_2}, then we obtain the set of
merge-legal edges depicted in Figure~\ref{ex_abstr_red_graph3_3}.
\end{Example}

The next theorem follows directly from the previous definition and
from the fact that Figure~\ref{ex1_flip_merge} contains the only
possible ways in which edges in merge-legal sets for $G$ can be
connected.

\begin{Theorem} 
Let $G \in \gredgr$, and denote by $\mathcal{F}$ be the group
generated by the flip operations w.r.t. $G$ under function
composition. Then each element of $\mathcal{F}$ is self-inverse,
thus $\mathcal{F}$ is Abelian, and $\mathcal{F}$ acts transitively
on $\legalmset_G$.
\end{Theorem}

Let $D = \{p_1, \ldots, p_l\} \subseteq \dom(G)$. Then we define
$\flipm_D = \flipm_{p_l} \ \cdots \ \flipm_{p_1}$. Since
$\mathcal{F}$ is Abelian, $\flipm_D$ is well defined. Moreover,
since each each element in $\mathcal{F}$ is self-inverse,
$\mathcal{F} = \{\flipm_D \mid D \subseteq \dom(G)\}$. Also, if
$D_1, D_2 \subseteq \dom(G)$ and $D_1 \not= D_2$, then
$\flipm_{D_1}(E) \not= \flipm_{D_2}(E)$. Thus the following holds.
\begin{Theorem} \label{th_char_set_legalmset}
Let $G \in \gredgr$. Then there is a bijection $Q: 2^{\dom(G)}
\rightarrow \mathcal{F}$ given by $Q(D) = \flipm_D$. Moreover, for
each $E \in \legalmset_G$, $\legalmset_G = \{\flipm_D(E) \mid D
\subseteq \dom(G)\}$.
\end{Theorem}

\section{Merging and Splitting Connected Components}
\label{sec_flip_merging_split} Let $G = B(E_1,E_2)$ be an abstract
reduction graph and let $E \in \legalmset_G$. In this section we
consider the effect of the flip operation on the pointer-component
graph defined on the abstract reduction graph $H = B(E_1,E)$. If we
are able to obtain, using flip operations, a pointer-component graph
consisting of one vertex, then $\slegalmset_G \not= \emptyset$, and
consequently by Theorem~\ref{th_iso_gredgr_redgr1}, $G$ is
isomorphic to a reduction graph.

However, first we need to define the notion of pointer-component
graph for abstract reduction graphs in general. Fortunately, this
generalization is trivial.
\begin{Definition}
Let $G \in \gredgr$. The \emph{pointer-component graph of $G$},
denoted by $\pcgr_G$, is a multigraph $(\zeta, E, \epsilon)$, where
$\zeta$ is the set of connected components of $G$, $E = \dom(G)$,
and $\epsilon$ is, for $e \in E$, defined by $\epsilon(e) = \{C \in
\zeta \mid C$ $\mbox{contains}$ $\mbox{vertices}$ $\mbox{labelled by
} e\}$.
\end{Definition}

\begin{figure}
$$ \xymatrix{
C_1 \ar@{-}[r]^5 & C_3 \ar@{-}[d]^9 & C_4 \ar@(u,l)@{-}[]_3 \ar@(u,r)@{-}[]^6 \ar@/^/@{-}[d]^7 \\
C_2 \ar@{-}[u]^2 \ar@{-}[ur]^4 & R & C_5 \ar@/^/@{-}[u]^8
} $$ \caption{The pointer-component graph of the abstract reduction
graph from Figure~\ref{ex_abstr_red_graph2_1}.}
\label{ex_abstr_red_graph2_2}
\end{figure}
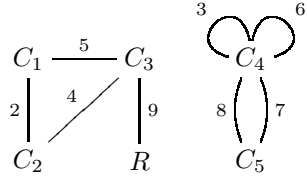

\begin{Example}
The pointer-component graph of the graph from
Figure~\ref{ex_abstr_red_graph2_1} is shown in
Figure~\ref{ex_abstr_red_graph2_2}.
\end{Example}

Note that when $G = B(E_1,E_2) \in \gredgr$ and $E \in
\legalmset_G$, then $E$ is desirable for $B$. Hence, $H = B(E_1,E)$
is also an abstract reduction graph. Therefore, e.g., $\pcgr_{H}$ is
defined.

It is useful to distinguish the pointers that form loops in the
pointer-component graph. Therefore, we define, for $G \in \gredgr$,
$\bridge(G) = \{e \in E \mid |\epsilon(e)| = 2\}$ where $\pcgr_G =
(V,E,\epsilon)$. In \cite{StrategiesSnr/Brijder06}, $\bridge(G)$ is
denoted as $\mathrm{snrdom}(G)$. However, this notation does not
make sense for its uses in this \paperchap.
\begin{Example}
From Figure~\ref{ex_abstr_red_graph2_2} it follows that $\bridge(G)
= \dom(G) \backslash \{3,6\}$ for the abstract reduction graph $G$
depicted in Figure~\ref{ex_abstr_red_graph2_1}.
\end{Example}

Merge rules have been used for multigraphs, and pointer-component
graphs in particular in \cite{StrategiesSnr/Brijder06}. The
definition presented here is slightly different from the one in
\cite{StrategiesSnr/Brijder06} -- here the pointer $p$ on which the
merge rule is applied remains present after the rule is applied.
\begin{Definition}
For each edge $p$, the \emph{$p$-merge rule}, denoted by $\merge_p$,
is a rule applicable to (defined on) multigraphs $G = (V, E,
\epsilon)$ with $p \in \bridge(G)$. It is defined by
$$
\merge_p(G) = (V', E, \epsilon'),
$$
where $V' = \left( V \backslash \epsilon(p) \right) \cup \{v'\}$
with $v' \not\in V$, and $\epsilon'(e) = \{h(v_1),h(v_2)\}$ iff
$\epsilon(e) = \{v_1,v_2\}$ where $h(v) = v'$ if $v \in
\epsilon(p)$, otherwise it is the identity.
\end{Definition}
It is easy to see that merge rules commute. We are now ready to
state the following result which is similar to Theorem~27 in
\cite{StrategiesSnr/Brijder06}.
\begin{Theorem} \label{th_merge_flip}
Let $G = B(E_1,E_2) \in \gredgr$, let $E \in \legalmset_G$, let $H =
B(E_1,E)$, and let, for $p \in \dom(G)$, $H_p =
B(E_1,\flipm_{p}(E))$.
\begin{itemize}
\item
If $p \in \bridge(H)$, then $\pcgr_{H_p} \approx
\merge_p(\pcgr_{H})$
\\
(and therefore $o(\pcgr_{H_p}) = o(\pcgr_{H}) - 1$).
\item
If $p \in \dom(H) \backslash \bridge(H)$, then $o(\pcgr_{H}) \leq
o(\pcgr_{H_p}) \leq o(\pcgr_{H})+1$.
\end{itemize}
\end{Theorem}
\begin{Proof}
First let $p \in \bridge(H)$. Let $\fiberonlabel_{H,p} =
\{\{v_1,v_2\},\{v_3,v_4\}\}$. Then, $H$ has the following form,
where each of the two edges in $\fiberonlabel_{H,p}$ are from
different connected components in $H$ and where, unlike our
convention, we have depicted the vertices by their identity instead
of their label:
$$
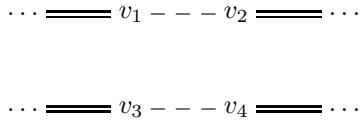
\xymatrix{
\ldots \redge[r] &
v_1 \medge[r] & v_2 \redge[r] &
\ldots \\
\ldots \redge[r] &
v_3  \medge[r] & v_4 \redge[r] &
\ldots
}
$$
Now, either $\{\{v_1,v_4\},\{v_2,v_3\}\} \subseteq E_2$ or
$\{\{v_1,v_3\},\{v_2,v_4\}\} \subseteq E_2$. Thus $H_p$ is of either

$$
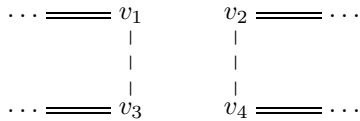
\xymatrix{
\ldots \redge[r] &
v_1 \medge[d] & \medge[d] v_2 \redge[r] &
\ldots \\
\ldots \redge[r] &
v_3  & v_4 \redge[r] &
\ldots
}
$$

or

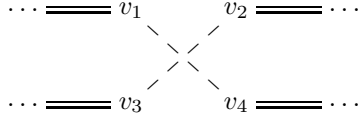
$$
\xymatrix{
\ldots \redge[r] &
v_1 \medge[dr] & \medge[dl] v_2 \redge[r] &
\ldots \\
\ldots \redge[r] &
v_3  & v_4 \redge[r] &
\ldots
}
$$

form, respectively. Thus in both cases, the two connected components
are merged, and thus $\pcgr_{H_p}$ can be obtained (up to
isomorphism) from $\pcgr_{H}$ by applying the $\merge_p$ operation.

Now let $p \in \dom(H) \backslash \bridge(H)$. Then the edges in
$\fiberonlabel_{H,p}$ belong to the same connected component. Thus
$H$ has the following form
$$
\xymatrix{
\cdots \redge[r] & v_1 \medge[r] & v_2 \redge[r] & \cdots \redge[r]
& v_3 \medge[r] & v_4 \redge[r] & \cdots
}
$$
where $\fiberonlabel_{H,p} = \{\{v_1,v_2\},\{v_3,v_4\}\}$. Again,
either $\{\{v_1,v_4\},\{v_2,v_3\}\} \subseteq E_2$ or
$\{\{v_1,v_3\},\{v_2,v_4\}\} \subseteq E_2$. Thus $H_p$ is of either

$$
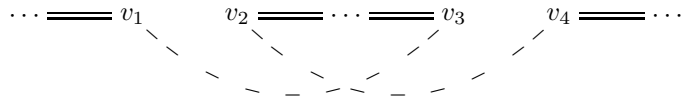
\xymatrix{
\cdots \redge[r] & v_1 \medge@/_2.5pc/[rrr] & v_2
\medge@/_2.5pc/[rrr] \redge[r] & \cdots \redge[r] & v_3 &
v_4 \redge[r] & \cdots \\
\\
}
$$

or

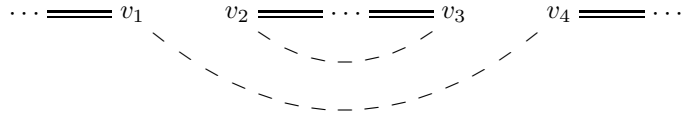
$$
\xymatrix{
\cdots \redge[r] & v_1 \medge@/_3pc/[rrrr] & v_2 \medge@/_1.5pc/[rr]
\redge[r] & \cdots \redge[r] & v_3 &
v_4 \redge[r] & \cdots \\
\\
}
$$

form, respectively. Thus, $H_p$ has either the same number of
connected components of $H$ or exactly one more, respectively. Thus,
$o(\pcgr_{H}) \leq o(\pcgr_{H_p}) \leq o(\pcgr_{H})+1$.
\end{Proof}

\begin{Example}
Let $G = B(E_1,E_2) \in \gredgr$ be as in
Figure~\ref{ex_abstr_red_graph3_1}. If we take $E \in \legalmset_G$
as in Figure~\ref{ex_abstr_red_graph3_2}, then $2 \in \bridge(H)$
with $H = B(E_1,E)$. Therefore, by Theorem~\ref{th_merge_flip} and
the fact that $G$ has exactly two connected components, $H_2 =
B(E_1,\flipm_{2}(E))$ is a connected graph. Indeed, this is clear
from Figure~\ref{ex_abstr_red_graph3_3} (by ignoring the edges from
$E_2$).
\end{Example}

Informally, the next lemma shows that by applying flip operations,
we can shrink a connected pointer-component graph to a single
vertex. In this way, the underlying abstract reduction graph is a
connected graph.

\begin{Remark}
The next lemma appears to be similar to Lemma~29 in
\cite{StrategiesSnr/Brijder06}. Although the flip operation (defined
on graphs) and the rem operation (defined on strings) are quite
distinct, they do have a similar effect on the pointer-component
graph.
\end{Remark}

\begin{Lemma} \label{lem_G_G'_conn_comp}
Let $G = B(E_1,E_2) \in \gredgr$, let $E \in \legalmset_G$, let $H =
B(E_1,E)$, and let $D \subseteq \dom(G) = \dom(H)$. Then
$\pcgr_{H}|_{D}$ is a tree iff $B(E_1,\flipm_D(E))$ and $H$ have $1$
and $|D|+1$ connected components, respectively.
\end{Lemma}
\begin{Proof}
Let $D = \{p_1, \ldots, p_n\}$. We first prove the forward
implication. If $\pcgr_{H}|_{D}$ is a tree, then it has $|D|$ edges,
and thus $|D|+1$ vertices. Therefore, $\pcgr_{H}$ has $|D|+1$
vertices, and consequently, $H$ has $|D|+1$ connected components.
Since $\pcgr_{H}|_{D}$ is acyclic, by Theorem~\ref{th_merge_flip},
$$
\pcgr_{B(E_1,\flipm_D(E))} = \pcgr_{B(E_1,(\flipm_{p_n} \ \cdots \
\flipm_{p_1})(E))} \approx (\merge_{p_n} \ \cdots \
\merge_{p_1})(\pcgr_{H}).
$$
Now, applying $|D|$ merge operations on a graph with $|D|+1$
vertices, results in a graph containing exactly one vertex. Thus
$B(E_1,\flipm_D(E))$ has one connected component.

We now prove the reverse implication. Moving from $H = B(E_1,E)$ to
$B(E_1,\flipm_D(E))$ reduces the number of connected components in
$|D|$ steps from $|D|+1$ to $1$. By Theorem~\ref{th_merge_flip},
each flip operation of $\flipm_D$ corresponds to a merge operation.
Therefore $(\merge_{p_n} \ \cdots \ \merge_{p_1})$ is applicable to
$\pcgr_{H}$. Consequently, $\pcgr_{H}|_{D}$ is acyclic. Since this
graph has $|D|+1$ vertices, $\pcgr_{H}|_{D}$ is a tree.
\end{Proof}

\section{Connectedness of Pointer-Component Graph}
\label{sec_conn_pc_graph} In this section we use the results of the
previous two sections to prove our first main result, cf.
Theorem~\ref{th_char_iso_redgr}, which strengthens
Theorem~\ref{th_iso_gredgr_redgr1} by replacing the requirement
$\slegalmset_G \not= \emptyset$ by a simple test on $\pcgr_G$.
%
%
%
We now characterize the connectedness of $\pcgr_G$.

\begin{Definition}
Let $B = (V,f,s,t)$ be a coloured base. We say that a set of edges
$E$ for $B$ is \emph{well-coloured (for $B$)} if for each partition
$\rho = (V_1,V_2)$ of $V$ with $f(V_1) \cap f(V_2) = \emptyset$,
there is an edge $\{v_1,v_2\} \in E$ with $v_1 \in V_1$ and $v_2 \in
V_2$.
\end{Definition}
We call $G = B(E_1,E_2) \in \gredgr$ \emph{well-coloured} if $E_1$
is well-coloured for $B$.

\begin{Lemma} \label{lem_G_G'_connected}
Let $G \in \gredgr$. Then $\pcgr_{G}$ is a connected graph iff $G$
is well-coloured.
\end{Lemma}
\begin{Proof}
Let $G = B(E_1,E_2)$ with $B = (V,f,s,t)$. We first prove the
forward implication. Let $G$ be not well-coloured. Then there is a
partition $\rho = (V_1,V_2)$ of $V$ with $f(V_1) \cap f(V_2) =
\emptyset$ such that for each $e \in E_1$, either $e \subseteq V_1$
or $e \subseteq V_2$. Since for each $\{v_1,v_2\} \in E_2$ we have
$f(v_1) = f(v_2)$, we have either $\{v_1,v_2\} \subseteq V_1$ or
$\{v_1,v_2\} \subseteq V_2$. Therefore $V_1$ and $V_2$ induce two
non-empty sets of connected components which have no vertex label in
common. Therefore, $\pcgr_{G}$ is not a connected graph.

We now prove the reverse implication. Assume that $\pcgr_{G} =
(\zeta,E,\epsilon)$ is not a connected graph. Then, by the
definition of pointer-component graph, there is a partition
$(C_1,C_2)$ of $\zeta$ such that $C_1$ and $C_2$ have no vertex
label in common. Let $V_i$ be the set of vertices of the connected
components in $C_i$ ($i \in \{1,2\}$). Then for partition $\rho =
(V_1,V_2)$ of $V$ we have $f(V_1) \cap f(V_2) = \emptyset$ and for
each $e \in E_1 \cup E_2$, either $e \subseteq V_1$ or $e \subseteq
V_2$. Therefore $G$ is not well-coloured.
\end{Proof}
Clearly, if $G = B(E_1,E_2) \in \gredgr$ is well-coloured and $E$ is
desirable for $B$ (e.g., one could take $E \in \legalmset_G$), then
$H = B(E_1,E) \in \gredgr$ and $H$ is well-coloured. Therefore, by
Lemma~\ref{lem_G_G'_connected}, $\pcgr_{G}$ is a connected graph iff
$\pcgr_{H}$ is a connected graph.

By Theorem~\ref{th_iso_gredgr_redgr1} the next result is essential
to efficiently determine which abstract reduction graphs are
isomorphic to reduction graphs.
\begin{Theorem} \label{th_G_conn_not_empty}
Let $G \in \gredgr$. Then $\pcgr_{G}$ is a connected graph iff
$\slegalmset_G \not= \emptyset$.
\end{Theorem}
\begin{Proof}
Let $G = B(E_1,E_2)$. We first prove the forward implication. Let
$\pcgr_{G}$ be a connected graph and let $E \in \legalmset_G$. Then
$\pcgr_{H}$ with $H = B(E_1,E)$ is a connected graph. Thus there
exists a $D \subseteq \dom(G)$ such that $\pcgr_{H}|_{D}$ is a tree.
By Lemma~\ref{lem_G_G'_conn_comp}, $B(E_1,\flipm_D(E))$ is a
connected graph, and consequently $\flipm_D(E) \in \slegalmset_G$.

We now prove the reverse implication. Let $E \in \slegalmset_G$.
Thus, $H = B(E_1,E)$ is a connected graph, and hence $\pcgr_{H}$ is
a connected graph. Therefore, $\pcgr_{G}$ is also a connected graph.
\end{Proof}

We can summarize the last two results as follows.
\begin{Corollary} \label{cor_equiv_well_col}
Let $G \in \gredgr$. Then the following conditions are equivalent:
\begin{enumerate}
\item $G$ is well-coloured,
\item $\pcgr_{G}$ is a connected graph, and
\item $\slegalmset_G \not= \emptyset$.
\end{enumerate}
\end{Corollary}

\begin{Example}
By Figure~\ref{ex_abstr_red_graph1_2} and
Corollary~\ref{cor_equiv_well_col}, for (abstract) reduction graph
$G_1$ in Figure~\ref{ex_abstr_red_graph1_1} we have
$\slegalmset_{G_1} \not= \emptyset$. On the other hand, by
Figure~\ref{ex_abstr_red_graph2_2} and
Corollary~\ref{cor_equiv_well_col}, for abstract reduction graph
$G_2$ in Figure~\ref{ex_abstr_red_graph2_1} we have
$\slegalmset_{G_2} = \emptyset$.
\end{Example}

By Corollary~\ref{cor_equiv_well_col} and
Theorem~\ref{th_iso_gredgr_redgr1} we obtain the first main result
of this \paperchap. It shows that one needs to check only a few
computationally easy conditions to determine whether or not a 2-edge
coloured graph is (isomorphic to) a reduction graph. Surprisingly,
the `high-level' notion of pointer-component graph is crucial in
this characterization.
\begin{Theorem} \label{th_char_iso_redgr}
Let $G$ be a 2-edge coloured graph. Then $G$ isomorphic to a
reduction graph iff $G \in \gredgr$ and $\pcgr_{G}$ is a connected
graph.
\end{Theorem}
Note that in the previous theorem we can equally well replace
``$\pcgr_{G}$ is a connected graph'' by one of the other equivalent
conditions in Corollary~\ref{cor_equiv_well_col}.

In Theorem~21 in \cite{StrategiesSnr/Brijder06} it is shown that the
pointer-component graph of each reduction graph is a connected
graph. We did not use that result here -- in fact it is now a direct
consequence of Theorem~\ref{th_char_iso_redgr}.

Not only is it computationally efficient to determine whether or not
a 2-edge coloured graph $G$ is isomorphic to a reduction graph, but,
when this is the case, then it is also computationally easy to
determine a legal string $u$ for which $G \approx \redgr{u}$.
Indeed, we can determine such a $u$ from $G = B(E_1,E_2)$ as
follows:
\begin{enumerate}
\item
Determine a $E \in \legalmset_G$. As we have mentioned before, such
a $E$ is easily obtained.
\item
Compute $\pcgr_{H}$ with $H = B(E_1,E)$, and determine a set of
edges $D$ such that $\pcgr_{H}|_D$ is a tree.
\item
Compute $G' = B(E_1,E_2,\flipm_{D}(E))$, and determine a $u \in
\setlegalgredgr_{G'}$.
\end{enumerate}

As a consequence, pointer-component graphs of legal strings can,
surprisingly, take all imaginable forms.
\begin{Corollary}
Every connected multigraph $G = (V,E,\epsilon)$ with $E \subseteq
\Delta$ is isomorphic to a pointer-component graph of a legal
string.
\end{Corollary}



\section{Flip and the Underlying Legal String} \label{sec_second_main_result1}
We now move to the second part of this \paperchap, where we
characterize the fibers $\mathcal{R}^{-1}(\redgr{u})$ modulo graph
isomorphism. First we consider the effect of flip operations on the
set of merge edges.
\begin{Lemma} \label{lem_flip_neg}
Let $u$ be a legal string and let $p \in \dom(u)$. If $p$ is
negative in $u$, then $\flipm_p(\mergeEdges_u) \in \slegalmset_u$.
If $p$ is positive in $u$, then $\flipm_p(\mergeEdges_u) \not\in
\slegalmset_u$. In other words, $\flipm_p(\mergeEdges_u) \in
\slegalmset_u$ iff $p$ is negative in $u$.
\end{Lemma}
\begin{Proof}
Let $\redgr{u} = B(E_1,E_2)$. By the definition of $\flipm_p$,
$\flipm_p(\mergeEdges_u) \in \legalmset_u$. It suffices to prove
that $G = B(E_1,\flipm_p(\mergeEdges_u))$ is a connected graph when
$p$ is negative in $u$ and not a connected graph when $p$ is
positive in $u$. Graph $B(E_1,\mergeEdges_u)$ has the following
form:
$$
\scalebox{0.72}{\xymatrix{
s \redge[r] & \pset{p}_1 \medge[r] & \pset{p}_1 \redge[r] & \cdots
\redge[r] & \pset{p} \medge[r] & \pset{p} \redge[r] & \cdots
\redge[r] & \pset{p} \medge[r] & \pset{p} \redge[r] & \cdots
\redge[r] & \pset{p}_n \medge[r] & \pset{p}_n \redge[r] & t
}}
$$
Now if $p$ is negative in $u$, then $G$ has the following form:
$$
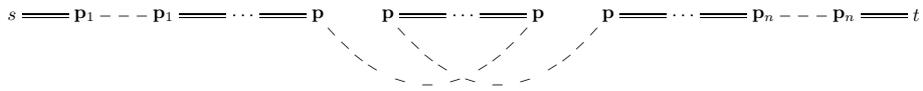
\scalebox{0.72}{\xymatrix{
s \redge[r] & \pset{p}_1 \medge[r] & \pset{p}_1 \redge[r] & \cdots
\redge[r] & \pset{p} \ar@{--}@/_3pc/[rrr] & \pset{p}
\ar@{--}@/_3pc/[rrr] \redge[r] & \cdots \redge[r] & \pset{p} &
\pset{p} \redge[r] & \cdots \redge[r] & \pset{p}_n \medge[r] &
\pset{p}_n \redge[r] & t \\
\\
}}
$$
Thus in this case $G$ is connected.

If $p$ is positive in $u$, then $G$ has the following form:
$$
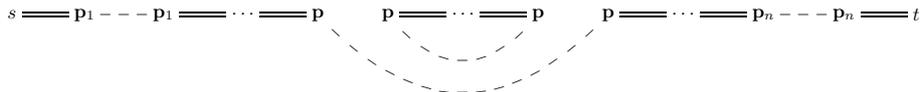
\scalebox{0.72}{\xymatrix{
s \redge[r] & \pset{p}_1 \medge[r] & \pset{p}_1 \redge[r] & \cdots
\redge[r] & \pset{p} \ar@{--}@/_3.4pc/[rrrr] & \pset{p}
\ar@{--}@/_2.0pc/[rr] \redge[r] & \cdots \redge[r] & \pset{p} &
\pset{p} \redge[r] & \cdots \redge[r] & \pset{p}_n \medge[r] &
\pset{p}_n \redge[r] & t \\
\\
}}
$$
Thus in this case $G$ is not connected.
\end{Proof}

\begin{Lemma} \label{lem_flip_double}
Let $u$ be a legal string and let $p,q \in \dom(u)$. If $p$ and $q$
are overlapping in $u$ \emph{and} not both negative in $u$, then
$\flipm_{\{p,q\}}(\mergeEdges_u) \in \slegalmset_u$.
\end{Lemma}
\begin{Proof}
Let $\redgr{u} = B(E_1,E_2)$. Then $B(E_1,\mergeEdges_u)$ has the
following form (we can assume without loss of generality that $p$
appears before $q$ in the path from $s$ to $t$):
$$
\scalebox{0.63}{\xymatrix{
s \redge[r] & \cdots \redge[r] &
\pset{p} \medge[r] & \pset{p} \redge[r] & \cdots \redge[r] &
\pset{q} \medge[r] & \pset{q} \redge[r] & \cdots \redge[r] &
\pset{p} \medge[r] & \pset{p} \redge[r] & \cdots \redge[r] &
\pset{q} \medge[r] & \pset{q} \redge[r] & \cdots \redge[r] & t
}}
$$
Assume that $p$ is positive in $u$ -- the other case ($q$ is
positive in $u$) is proved similarly. By the proof of
Lemma~\ref{lem_flip_neg} it follows that
$B(E_1,\flipm_p(\mergeEdges_u))$ has the following form:
$$
\scalebox{0.63}{\xymatrix{
s \redge[r] & \cdots \redge[r] &
\pset{p} \ar@{--}@/_4.5pc/[rrrrrrr] & \pset{p}
\ar@{--}@/_3pc/[rrrrr] \redge[r] & \cdots \redge[r] &
\pset{q} \medge[r] & \pset{q} \redge[r] & \cdots \redge[r] &
\pset{p} & \pset{p} \redge[r] & \cdots \redge[r] &
\pset{q} \medge[r] & \pset{q} \redge[r] & \cdots \redge[r] & t\\
\\
\\
}}
$$
Therefore, $q \in \bridge(B(E_1,\flipm_p(\mergeEdges_u)))$. By
Theorem~\ref{th_merge_flip}, the pointer-component graph of
$B(E_1,\flipm_{\{p,q\}}(\mergeEdges_u))$ has only one vertex.
Consequently, $B(E_1,\flipm_{\{p,q\}}(\mergeEdges_u))$ is connected
and thus $\flipm_{\{p,q\}}(\mergeEdges_u) \in \slegalmset_u$.
\end{Proof}

\begin{Lemma} \label{lem_flip_applicable}
Let $u$ be a legal string, and let $D \subseteq \dom(u)$ be
nonempty. If $\flipm_D(\mergeEdges_u) \in \slegalmset_u$, then
either there is a $p \in D$ negative in $u$ or there are $p,q \in D$
positive and overlapping in $u$.
\end{Lemma}
\begin{Proof}
Let $\eredgr{u} = B(E_1,E_2,\mergeEdges_u)$ and let
$\flipm_D(\mergeEdges_u) \in \slegalmset_u$. Then
$B(E_1,\flipm_D(\mergeEdges_u))$ is a connected graph. Assume to the
contrary that all elements in $D$ are positive and pairwise
non-overlapping in $u$. Then there is a $p \in D$ such that the
domain of the $p$-interval does not contain an element in $D
\backslash \{p\}$. By the proof of Lemma~\ref{lem_flip_neg}
$B(E_1,\flipm_p(\mergeEdges_u))$ consist of two connected
components, one of which does not have vertices labelled by elements
in $D \backslash \{p\}$. Therefore $B(E_1,\flipm_D(\mergeEdges_u))$
also contains this connected component, and thus
$B(E_1,\flipm_D(\mergeEdges_u))$ has more than one connected
component -- a contradiction.
\end{Proof}

By the previous lemmata, we have the following result.
\begin{Theorem} \label{th_flip_applicable}
Let $u$ be a legal string, and let $D \subseteq \dom(u)$ be
nonempty. If $\flipm_D(\mergeEdges_u) \in \slegalmset_u$, then
either there is a $p \in D$ negative in $u$ with
$\flipm_p(\mergeEdges_u) \in \slegalmset_u$ or there are $p,q \in D$
positive and overlapping in $u$ with
$\flipm_{\{p,q\}}(\mergeEdges_u) \in \slegalmset_u$.
\end{Theorem}

\section{Dual String Rules} \label{sec_second_main_result2}
We now define the dual string rules. These rules will be used to
characterize the effect of flip operations on the underlying legal
string. For all $p,q \in \Pi$ with $\pset{p} \not = \pset{q}$ we
define
\begin{itemize}
\item
the \emph{dual string positive rule} for $p$ is defined by
$\textbf{dspr}_{p}(u_1 p u_2 p u_3) = u_1 p \bar u_2 p u_3$,
\item
the \emph{dual string double rule} for $p,q$ is defined by
$\textbf{dsdr}_{p,q}(u_1 p u_2 q u_3 \bar p u_4 \bar q u_5) = u_1 p
u_4 q u_3 \bar p u_2 \bar q u_5$,
\end{itemize}
where $u_1,u_2,\ldots,u_5$ are arbitrary (possibly empty) strings
over $\Pi$. Notice that the dual string rules are self-inverse. Also
notice the strong similarities between $\textbf{dspr}$ and
$\textbf{spr}$, and between $\textbf{dsdr}$ and $\textbf{sdr}$. Both
$\textbf{dspr}_p$ and $\textbf{spr}_p$ invert the substring between
the two occurrences of $p$ or $\bar p$. However, $\textbf{dspr}_p$
is applicable when $p$ is negative, while $\textbf{spr}_p$ is
applicable when $p$ is positive. Also, $\textbf{spr}_p$ removes the
occurrences of $p$ and $\bar p$, while $\textbf{dspr}$ does not. A
similar comparison can be made between $\textbf{dsdr}$ and
$\textbf{sdr}$.

The domain of (sequences of) dual string rules is defined similarly
as for string rules. Thus, e.g., $\dom(\textbf{dsdr}_{p,q}) =
\{p,q\}$.
%

\begin{Definition}
Let $u$ and $v$ be legal strings. We say that $u$ and $v$ are
\emph{dual}, denoted by $\approx_d$ if there is a (possibly empty)
sequence $\varphi$ of dual string rules applicable to $u$ such that
$\varphi(u) \approx v$.
\end{Definition}
Notice that $\approx_d$ is an equivalence relation. Clearly,
$\approx_d$ is reflexive. It is symmetrical since dual string rules
are self-inverse, and it is transitive by function composition: if
$\varphi_1(u) \approx v$ and $\varphi_2(v) \approx w$, then
$(\varphi_2 \ \varphi_1)(u) \approx w$.

Since $\textbf{dspr}_p$ is applicable when $p$ is negative in $u$
and $\textbf{dsdr}_{p,q}$ is applicable when $p$ and $q$ are
positive and overlapping, the following result is a direct corollary
to Lemma~\ref{lem_flip_applicable}.
\begin{Corollary} \label{cor_dual_applicable}
Let $u$ be a legal string, and let $D \subseteq \dom(u)$ be
nonempty. If $\flipm_D(\mergeEdges_u) \in \slegalmset_u$, then there
is a dual string rule $\rho$ with $\dom(\rho) \subseteq D$
applicable to $u$.
\end{Corollary}

Let $\varphi = \rho_n \ \cdots \ \rho_1$ with each $\rho_i$ (for $1
\leq i \leq n$) a dual string rule. We define $\odom(\varphi) =
\bigoplus_{1 \leq i \leq n} \dom(\rho_i)$. Thus, $\odom(\varphi)
\subseteq \dom(\varphi)$. We call $\varphi$ \emph{reduced} if
$\dom(\rho_i) \cap \dom(\rho_j) = \emptyset$ for all $1 \leq i < j
\leq n$. Note that if $\varphi$ is reduced, then $\dom(\varphi) =
\odom(\varphi)$.

Let $G = B(E_1,E_2,E_3)$ be an extended abstract reduction graph,
and let $D \subseteq \dom(G)$. Then we define $\flipm_{D}(G) =
B(E_1,E_2,\flipm_{G',D}(E_3))$, where $G' = B(E_1,E_2)$.


\begin{Lemma} \label{lem_dual_ops_appl}
Let $u$ be a legal string, and let $\varphi$ be a sequence of dual
string rules applicable to $u$. Then $\eredgr{\varphi(u)} \approx
\flipm_{D}(\eredgr{u})$ with $D = \odom(\varphi)$. Consequently,
$\redgr{\varphi(u)} \approx \redgr{u}$.
\end{Lemma}
\begin{Proof}
It suffices to prove the result for the case $\varphi =
\textbf{dspr}_{p}$ with $p \in \Pi$ and for the case $\varphi =
\textbf{dsdr}_{p,q}$ with $p,q \in \Pi$. We first prove the case
where $\varphi = \textbf{dspr}_{p}$ for some $p \in \Pi$ is
applicable to $u$. Then by the second figure in the proof of
Lemma~\ref{lem_flip_neg} we see that the inversion of the substring
between the two occurrences of $p$ in $u$ accomplished by $\varphi$
faithfully simulates the corresponding effect of $\flipm_{p}$ on
$\eredgr{u}$. We only need to verify that $p$ is negative in
$\flipm_{p}(\eredgr{u})$. To do this, we depict $\eredgr{u}$ such
that the vertices are represented by their identity instead of their
label:
$$
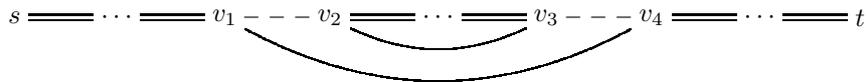
\xymatrix{
s \redge[r] & \cdots \redge[r] & v_1 \medge[r] \dedge@/_2pc/[rrrr] &
v_2 \dedge@/_1.0pc/[rr] \redge[r] & \cdots \redge[r] & v_3
\medge[r] & v_4 \redge[r] & \cdots \redge[r] & t \\
\\
}
$$
where the vertices $v_i$, $i \in \{1,2,3,4\}$, are labelled by
$\pset{p}$. Then $\flipm_{p}(\eredgr{u})$ is
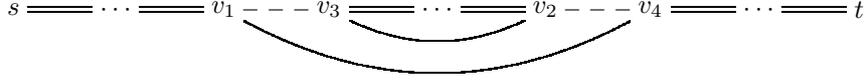
$$
\xymatrix{
s \redge[r] & \cdots \redge[r] & v_1 \medge[r] \dedge@/_2pc/[rrrr] &
v_3 \dedge@/_1.0pc/[rr] \redge[r] & \cdots \redge[r] & v_2
\medge[r] & v_4 \redge[r] & \cdots \redge[r] & t \\
\\
}
$$
Therefore $p$ is indeed negative in $\flipm_{p}(\eredgr{u})$, and
consequently $\eredgr{\varphi(u)} \approx \flipm_{p}(\eredgr{u})$.

We now prove the case where $\varphi = \textbf{dsdr}_{p,q}$ with
$p,q \in \Pi$. Let $\eredgr{u} = B(E_1,E_2,E_3)$, then $\eredgr{u}$
has the following form
$$
\scalebox{0.63}{\xymatrix{
s \redge[r] & \cdots \redge[r] &
\pset{p} \medge[r] & \pset{p} \redge[r] & \cdots \redge[r] &
\pset{q} \medge[r] & \pset{q} \redge[r] & \cdots \redge[r] &
\pset{p} \medge[r] & \pset{p} \redge[r] & \cdots \redge[r] &
\pset{q} \medge[r] & \pset{q} \redge[r] & \cdots \redge[r] & t
}}
$$
where we omitted the edges in $E_2$. Since $p$ and $q$ are positive
in $u$, $\flipm_{\{p,q\}}(\eredgr{u})$ has the following form:
$$
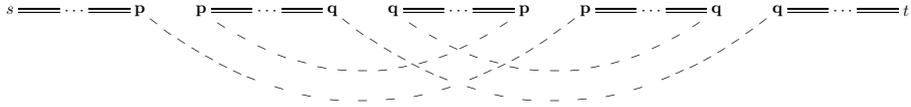
\scalebox{0.63}{\xymatrix{
s \redge[r] & \cdots \redge[r] &
\pset{p} \medge@/_4.5pc/[rrrrrrr] & \pset{p} \ar@{--}@/_3pc/[rrrrr]
\redge[r] & \cdots \redge[r] &
\pset{q} \medge@/_4.5pc/[rrrrrrr] & \pset{q} \ar@{--}@/_3pc/[rrrrr]
\redge[r] & \cdots \redge[r] &
\pset{p} & \pset{p} \redge[r] & \cdots \redge[r] &
\pset{q} & \pset{q} \redge[r] & \cdots \redge[r] & t\\
\\
\\
}}
$$
where we again omitted the edges in $E_2$. Thus, we see that
interchanging the substring in $u$ between $p$ and $q$ and the
substring in $u$ between $\bar p$ and $\bar q$ accomplished by
$\varphi$ faithfully simulates the corresponding effect of
$\flipm_{p,q}$ on $\eredgr{u}$. We only need to verify that both $p$
and $q$ are positive in $\flipm_{p,q}(\eredgr{u})$. To do this, we
depict $\eredgr{u}$ such that the vertices are represented by their
identity instead of their label:
$$
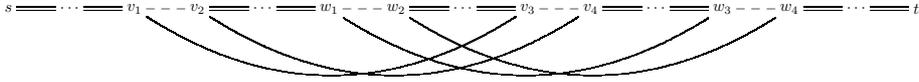
\scalebox{0.6}{\xymatrix{
s \redge[r] & \cdots \redge[r] &
v_1 \medge[r] \dedge@/_3.5pc/[rrrrrr] & v_2 \redge[r]
\dedge@/_3.5pc/[rrrrrr] & \cdots \redge[r] &
w_1 \medge[r] \dedge@/_3.5pc/[rrrrrr] & w_2 \redge[r]
\dedge@/_3.5pc/[rrrrrr] & \cdots \redge[r] &
v_3 \medge[r] & v_4 \redge[r] & \cdots \redge[r] &
w_3 \medge[r] & w_4 \redge[r] & \cdots \redge[r] & t
\\
\\
}}
$$
where the vertices $v_i$ and $w_i$, $i \in \{1,2,3,4\}$, are
labelled by $\pset{p}$ and $\pset{q}$, respectively. Then
$\flipm_{p,q}(\eredgr{u})$ is
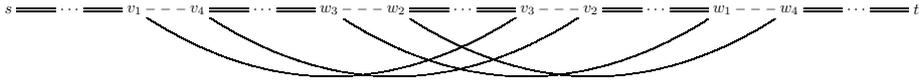
$$
\scalebox{0.6}{\xymatrix{
s \redge[r] & \cdots \redge[r] &
v_1 \medge[r] \dedge@/_3.5pc/[rrrrrr] & v_4 \redge[r]
\dedge@/_3.5pc/[rrrrrr] & \cdots \redge[r] &
w_3 \medge[r] \dedge@/_3.5pc/[rrrrrr] & w_2 \redge[r]
\dedge@/_3.5pc/[rrrrrr] & \cdots \redge[r] &
v_3 \medge[r] & v_2 \redge[r] & \cdots \redge[r] &
w_1 \medge[r] & w_4 \redge[r] & \cdots \redge[r] & t
\\
\\
}}
$$
Therefore both $p$ and $q$ are indeed positive in
$\flipm_{p,q}(\eredgr{u})$, and consequently $\eredgr{\varphi(u)}
\approx \flipm_{p,q}(\eredgr{u})$.
\end{Proof}
Thus, if $\varphi_1$ and $\varphi_2$ are sequences of dual string
rules applicable to a legal string $u$ with $\odom(\varphi_1) =
\odom(\varphi_2)$, then $\eredgr{\varphi_1(u)} \approx
\eredgr{\varphi_2(u)}$ and thus $\varphi_1(u) \approx \varphi_2(u)$.


\begin{Lemma} \label{lem_it_dual_op}
Let $u$ be a legal string, and let $D \subseteq \dom(u)$. There is a
reduced sequence $\varphi$ of dual string rules applicable to $u$
such that $\dom(\varphi) = D$ iff $\flipm_D(\mergeEdges_u) \in
\slegalmset_u$.
\end{Lemma}
\begin{Proof}
The forward implication follows directly from
Lemma~\ref{lem_dual_ops_appl}. We now prove the reverse implication.
If $D = \emptyset$, we have nothing to prove. Let $D \not=
\emptyset$. By Corollary~\ref{cor_dual_applicable}, there is a dual
string rule $\rho_1$ with $D_1 = \dom(\rho_1) \subseteq D$
applicable to $u$. By Lemma~\ref{lem_dual_ops_appl},
$\eredgr{\rho_1(u)} \approx \flipm_{D_1}(\eredgr{u})$ and $D_1 =
\odom(\rho_1) = \dom(\rho_1)$. Thus, $\flipm_{D \backslash D_1}
(\mergeEdges_{\rho_1(u)}) \in \slegalmset_{\rho_1(u)}$. Now by
iteration, there is a reduced sequence $\varphi$ of dual string
rules applicable to $u$ such that $\odom(\varphi) = \dom(\varphi) =
D$.
\end{Proof}
It follows from Lemma~\ref{lem_dual_ops_appl} and
Lemma~\ref{lem_it_dual_op} that reduced sequences of dual string
rules are a normal form of sequences of dual string rules. Indeed,
by Lemma~\ref{lem_dual_ops_appl}, if $\varphi$ is a sequence of dual
string rules applicable to a legal string $u$ with $D =
\odom(\varphi)$, then $\flipm_D(\mergeEdges_u) \in \slegalmset_u$.
By Lemma~\ref{lem_it_dual_op}, there is a reduced sequence
$\varphi'$ of dual string rules applicable to $u$ such that
$\dom(\varphi') = \odom(\varphi') = D$. By the paragraph below
Lemma~\ref{lem_dual_ops_appl}, we have $\varphi(u) \approx
\varphi'(u)$.

We are now ready to prove the second (and final) main result of this
\paperchap. It shows that $\mathcal{R}^{-1}(\redgr{u})$ (modulo
graph isomorphism) is the `orbit' of $u$ under the dual string
rules. That is, the legal strings obtained from $u$ by applying dual
string rules are exactly those legal strings to have the same
reduction graph as $u$ (up to isomorphism).
\begin{Theorem} \label{th_main_result}
Let $u$ and $v$ be legal strings. Then $u \approx_d v$ iff
$\redgr{u} \approx \redgr{v}$.
\end{Theorem}
\begin{Proof}
The forward implication follows directly from
Lemma~\ref{lem_dual_ops_appl}. We now prove the reverse implication.
Let $\redgr{u} \approx \redgr{v}$. By
Corollary~\ref{cor_iso_redgraph_merge}, there is a $E \in
\slegalmset_u$ such that $\eredgr{v} \approx B(E_1,E_2,E)$ with
$\redgr{u} = B(E_1,E_2)$. By Theorem~\ref{th_char_set_legalmset}, $E
= \flipm_D(\mergeEdges_u)$ for some $D \subseteq \dom(u)$. Since
$\flipm_D(\mergeEdges_u) \in \slegalmset_u$, by
Lemma~\ref{lem_it_dual_op}, there is a reduced sequence $\varphi$ of
dual string rules applicable to $u$ such that $\dom(\varphi) = D$.
Now by Lemma~\ref{lem_dual_ops_appl}, $\eredgr{\varphi(u)} \approx
\flipm_D(\eredgr{u}) \approx \eredgr{v}$, and therefore, by
Theorem~\ref{th_legal_unique_ext}, $\varphi(u) \approx v$.
\end{Proof}

\section{Discussion} \label{sec_discussion}
This \paperchap characterizes, having $\mathcal{R}$ as the function
which assigns to each legal string $u$ its reduction graph
$\redgr{u}$, the range of $\mathcal{R}$
(Theorem~\ref{th_char_iso_redgr}) and each fiber
$\mathcal{R}^{-1}(\redgr{u})$ modulo graph isomorphism
(Theorem~\ref{th_main_result}).

The first characterization corresponds to a computationally
efficient algorithm that determines whether or not a graph $G$ is
isomorphic to a reduction graph. Moreover, if this is the case, then
the algorithm given below Theorem~\ref{th_char_iso_redgr} allows for
an efficient determination of a legal string $u$ such that $G
\approx \redgr{u}$. The first characterization relies on the notion
of merge-legal edges and its flip operation introduced in this
\paperchap. In particular, the connected components in the subgraph
induced by the reality edges and the merge-legal edges and the flip
operation turns out to be relevant in this context.

The second characterization determines, given $u$, the whole set
$\mathcal{R}^{-1}(\redgr{u})$ modulo graph isomorphism. From a
biological point of view, the fibers characterize which micronuclear
genes obtain the same macronuclear structure. It turns out that
$\mathcal{R}^{-1}(\redgr{u})$ is the orbit of $u$ under the dual
string rules. Surprisingly, these two types of string rewriting
rules are very similar to the string positive rules and the string
double rules that are used to define the model. Moreover, each two
legal strings $u$ and $v$ in such a fiber can be transformed into
each other by a sequence $\varphi$ of string rewriting rules without
using a pointer more than once. Therefore, the number of string
rewriting rules in $\varphi$ can be bounded by the size of the
domain of $u$ (and $v$).

The reduction graph of a legal string $u$ in a certain sense retains
all information regarding applicability of string negative rules in
each successful reduction of $u$, while discarding almost all other
information regarding the rules applied in successful reductions,
see \cite{StrategiesSnr/Brijder06}. Therefore, the fiber in a sense
characterizes all legal strings that have the same properties
regarding the application of string negative rules. In biological
terms, this may allow for a way to determine whether or not the
strategies regarding the string negative rule are different among
the different kinds of (genes in) ciliates.

\paper{
\bibliography{../geneassembly}

\end{document}
}